\documentclass[preprint, 10pt, sort&compress]{elsarticle}

\usepackage[margin=2.5cm]{geometry}
\usepackage{setspace}
\usepackage{lmodern}
\usepackage{amsmath,amssymb}
\usepackage{amsfonts,amsthm}
\usepackage{lineno}
\usepackage{graphicx}
\usepackage{bm, bbm}
\usepackage{algorithm}
\usepackage{physics}
\usepackage{color}
\usepackage{pxfonts}
\usepackage[footnotesize,bf]{caption}
\usepackage{subcaption}
\usepackage{mathtools}
\usepackage{hhline}
\usepackage[T1]{fontenc}

\usepackage{caption}
\usepackage{graphicx}
\usepackage{float}
\usepackage{subcaption}

\usepackage{placeins}

\newtheorem{remark}{Remark}[section]

\makeatletter
\def\ps@pprintTitle{%
	\let\@oddhead\@empty
	\let\@evenhead\@empty
	\let\@oddfoot\@empty
	\let\@evenfoot\@oddfoot
}
\makeatother

\graphicspath{{figures/}}
\usepackage{multirow}
\usepackage{ulem}
\usepackage{tikz}
\usepackage{color}
\usepackage[]{xcolor}

\numberwithin{equation}{section} 

\begin{document}
\begin{frontmatter}
\title{Deep learning for the semi-classical limit of the Schr$\ddot{\text{o}}$dinger equation}
\author[a,b]{Jizu Huang}
 \author[a,b]{Rukang You}
  \author[a]{Tao Zhou}
\date{}
\address[a]{SKLMS, Academy of Mathematics and Systems Science, Chinese Academy
of Sciences, Beijing, 100190, PR China.}
\address[b]{School of Mathematical Sciences, University of Chinese Academy of Sciences, Beijing 100190, PR China.}

\cortext[cor]{Corresponding Author}

\begin{abstract}
In this paper, we integrate neural 
networks and Gaussian wave packets to numerically solve the Schr{\"o}dinger 
equation with a smooth potential near the semi-classical 
limit. Our focus is not only on accurately obtaining
solutions when the non-dimensional Planck's constant, 
$\varepsilon$, is small, but also on constructing an operator that maps initial values to solutions for the 
Schr{\"o}dinger equation with multiscale properties.
Using Gaussian wave packets framework, we first reformulate the Schrödinger equation as a system of ordinary differential equations. For a single initial condition, we solve the resulting system using PINNs or MscaleDNNs. Numerical simulations indicate that 
MscaleDNNs outperform PINNs, improving accuracy by one to two orders of magnitude. When dealing with a set of initial conditions, we adopt an operator-learning approach, such as physics-informed DeepONets. Numerical examples 
validate the effectiveness of physics-informed DeepONets with Gaussian wave packets in accurately mapping initial conditions to solutions.

\end{abstract}

\end{frontmatter}

\section{Introduction}
The semi-classical limit of the Schr{\"o}dinger equation 
presents significant computational challenges and 
holds numerous applications in chemistry and physics \cite{guillot1988semi, cingolani2002semiclassical, garashchuk20116, lasser2020computing}. 
Specifically, it involves the efficient computation of 
the following initial value problem:
\begin{flalign}
\left\{
\begin{aligned}
  \bm{\psi}_{t} &= \frac{\mathrm{i} \varepsilon}{2} \Delta\bm{\psi} - \frac{\mathrm{i}}{\varepsilon}\bm{V}(\bm{x}) \bm{\psi}, \\
\bm{\psi}(\bm{x},0)&=\varphi(\bm{x})\exp(\mathrm{i}\phi(\bm{x})/\varepsilon), 
\end{aligned}
\right.
\label{semi_sch} 
\end{flalign} 
where $\bm{\psi}=\bm{\psi}(\bm{x},t) \in \mathbb{C}, \ \varphi(\bm{x}) \ \text{and} \ \phi(\bm{x})$ 
are given smooth functions. Here, $\varepsilon$ denotes the non-dimensional Planck's 
constant, $\bm{V}(\bm{x})$ represents the potential. 
Our focus is on the semi-classical limit situations, where $\varepsilon  \ll  1 $. 
In this context, the solutions $\bm{\psi}(\bm{x},t)$ 
exhibit high-frequency oscillations in both space 
and time on a scale of $O (\varepsilon^{-1})$. \par 

When $\varepsilon$ is not exceedingly small, 
several accurate and efficient methods are available, particularly 
those based on operator splitting.
A classical approach utilizes Strang splitting 
(see, for example, \cite{bao2002time, pathria1990pseudo}), which 
achieves spectral  accuracy in $\Delta x/\varepsilon$ and a global error of $O(\Delta t^2/\varepsilon)$ 
in time. Another highly accurate splitting method, 
developed by Chin and Chen \cite{descombes2010exact}, 
attains spectral accuracy in space and fourth-order accuracy in time. However, since the global truncate error is scaled by a factor $1/\varepsilon$, the spatial mesh size and the time step size must be reduced synchronously with the decrease of $\varepsilon$. This makes these methods inefficient for solving semi-classical limit of the Schr{\"o}dinger equation when $\varepsilon\ll 1$. 
To overcome this limitation, 
two powerful approaches are commonly employed. 
The first is based on the Wentzel-Kramers-Brillouin (WKB) 
approximation, which seeks solutions 
of the form $\varphi(\bm{x},t)\exp(\mathrm{i}\phi(\bm{x},t)/\varepsilon)$ 
and derives partial differential equations 
for $\varphi$ and $\phi$. The second approach utilizes 
Gaussian beams or wave packets. A Gaussian wave packet is 
an exact solution of the Schr{\"o}dinger equation
for a harmonic potential and serves as 
an excellent approximation for 
an arbitrary smooth potential 
as $\varepsilon \to 0$; see, for example, Refs. 
\cite{faou2006poisson, hagedorn1980semiclassical, heller1975time, jin2008gaussian}. 
The advantage of this method is that it avoids 
the singularities encountered in the WKB approach. However, a 
drawback is that it is a Lagrangian method, and Gaussian beams 
may experience spreading or focusing over time.\par 

In recent years, the explosive growth of available data and computational 
resources has facilitated the successful application 
of Deep Neural Networks (DNNs) across a wide range of domains, including 
recommendation systems, speech recognition, mathematical physics, 
computer vision, pattern recognition, and more 
\cite{lecun2015deep, bishop2006pattern, li2020solving, krizhevsky2012imagenet, lake2015human}.
DNNs have also become a powerful tool for addressing forward and inverse problems in scientific computing \cite{raissi2019physics, yu2018deep, zang2020weak, ming2021deep, han2017deep, han2018deep, han2018solving, he2020relu, strofer2019data, wang2020mesh}. 
A prominent framework in this context is the Physics-Informed 
Neural Networks (PINNs) \cite{raissi2019physics}. 
However, due to their inherent tendency to preferentially learn low-frequency components, traditional DNNs perform poorly in solving high-frequency problems
\cite{rahaman2019spectral, xu2020frequency, zhang2019explicitizing, xu2018understanding}.
To address this challenge, multi-scale Deep Neural Networks (MscaleDNNs) \cite{liu2020multi} 
provide a significant advantage by enhancing the network's ability to capture high-frequency information through input scaling transformations. The MscaleDNNs have achieved great success in solving various equations \cite{cai2019multi, wang2020multi, li2020multi, li2023subspace, huang2025frequency}.
However, since the parameter $\varepsilon$ in 
the Schrödinger equation (\ref{semi_sch}) introduces high-frequency oscillations in both space and time on a scale of $O(\varepsilon^{-1})$, both PINNs and MscaleDNNs exhibit low accuracy in solving the Schrödinger equation when $\varepsilon\ll 1$. 
\par  

To alleviate these multi-scale difficulties, many deep learning methods leverage classical approaches, such as macro-micro or even-odd decomposition, as discussed in \cite{jin2023asymptotic, jin2024asymptotic, jin2024asymptoticv}. In this paper, we integrate the DNNs with Gaussian wave packets, leveraging the strengths of both approaches to solve the Schrödinger equation. By following the Gaussian wave packets framework, we transform the original Schrödinger equation (\ref{semi_sch}) into a system of ordinary differential equations (ODEs), which are subsequently solved using DNNs. We compare the performance of PINNs and MscaleDNNs in solving these ODEs. We evaluate and compare the performance of PINNs and MscaleDNNs in solving these ODEs. Due to the slow decay of the Fourier coefficients of the solutions to these ODEs, MscaleDNNs  exhibit significant advantages over PINNs, particularly in scenarios where $\varepsilon\ll 1$. In the semi-classical limit of the Schrödinger equation, 
different physical scenarios usually correspond to different initial conditions. The ability to rapidly obtain solutions for specific initial conditions is highly valuable.
Traditional methods, such as those presented in \cite{russo2013gaussian, faou2009computing}, rely on separate solvers to solve the Schrödinger equation (\ref{semi_sch}) for different initial conditions.
In contrast, neural networks introduce a paradigm shift through
operator learning, which maps infinite-dimensional 
input functions (the initial conditions) to infinite-dimensional output functions (the solutions). 
However, due to multi-scale challenges, operator learning also requires leveraging traditional methods \cite{wu2024capturing}. In this study, we employ Physics-Informed Deep Operator Networks (DeepONets) \cite{2021LuLu} to construct a mapping between the initial conditions and the solutions of the ODEs obtained derived from Gaussian wave packets. By leveraging the advantages of operator learning, we can efficiently obtain a series of solutions for various initial conditions.

\par 

This paper is structured as follows:  
\begin{itemize}
	\item \textbf{Section \ref{nn_semi_sch}} introduces the 
	fundamental concepts of Gaussian wave packets, PINNs,  
	MscaleDNNs, along with the 
	methodology for integrating these approaches to solve the Schrödinger equation (\ref{semi_sch}). 
	\item \textbf{Section \ref{sec3}} describes the application 
	of physics-informed DeepONets to tackle the operator problem of mapping initial values to solutions for the Schrödinger equation (\ref{semi_sch}) using Gaussian wave packets.
	\item \textbf{Section \ref{num_exp}} presents numerical experiments and results obtained from PINNs and MscaleDNNs with or without Gaussian wave packets.
	\item \textbf{Section \ref{sec5}} demonstrates operator learning in various dimensions, validating the feasibility of initial value operator mapping.
	\item \textbf{Section \ref{summary}} concludes the paper. 
\end{itemize}

\section{Neural networks for the semi-classical limit of the Schr{\"o}dinger equation}
\label{nn_semi_sch}
In this section, we first outline the application of PINNs
to solve the semi-classical 
limit of the Schr{\"o}dinger equation (\ref{semi_sch}). The presence of the small parameter $\varepsilon$ poses significant challenges for conventional DNNs, making them unsuitable for tackling this multi-scale problem effectively. 
To address this, in subsection \ref{pinn_gwp}, we introduce 
Gaussian wave packets to reformulate the Schr{\"o}dinger equation
into an ODE system, and explore  
PINNs to solve the resulting ODE system. However, the accuracy of PINNs remains inadequate for the ODE system derived from Gaussian wave packets. In the subsection \ref{ms_gwp},
we introduce MscaleDNNs 
and demonstrate their superior performance when combined with Gaussian wave packets. 

\subsection{PINNs for the semi-classical limit of the Schr{\"o}dinger equation}
\label{intro_pinn}
We begin by reviewing PINNs and their direct application to 
the semi-classical limit of the Schr{\"o}dinger 
equation (\ref{semi_sch}).
Let $\Omega\subset \mathbb{R}^d$ be a bounded spatial domain and 
$(\bm{x}, t) \in \Omega \times [0, T]$.
Let us consider periodic boundary conditions 
characterized by the periodic operator $\mathcal{P}$ in $d$-dimensions, 
defined as 
\begin{equation*}
	\mathcal{P}(\psi(\bm{x},t))=\psi(\bm{x}+\bm{L}, t),
\end{equation*}
where $\bm{L}=(L_1,L_2,\dots,L_d)$ represents the 
periodicity. PINNs employ a DDNs function $\psi(\bm{x},t;\theta)$ 
with adjustable parameters $\theta$ to approximate the unknown 
solution $\psi(\bm{x},t)$.
The optimal parameters are obtained by solving the 
following soft-constrained  optimization problem:
\begin{equation}
	\label{loss_sch}
	\mathop{\text{min}}\limits_{\theta \in \Theta} \mathcal{L}(\theta)=\mathop{\text{min}}\limits_{\theta \in \Theta}\left\{ \  w_r \cdot \mathcal{L}_r(\theta) + w_b \cdot \mathcal{L}_b(\theta)+ w_i \cdot \mathcal{L}_i(\theta)\right\},
\end{equation}
where $\Theta$ is the parameter space and 
$\{w_r, \ w_b, \ w_i\}$ are weights 
to balances the PDE loss $\mathcal{L}_r(\theta)$, the boundary 
loss $\mathcal{L}_b(\theta)$, and the initial loss $\mathcal{L}_i(\theta)$.
A common choice for loss $\mathcal{L}(\theta)$ is the $L^2$ loss, i.e.,
\begin{equation*}
	\begin{split}
		\mathcal{L}_r(\theta) &= \left\|\psi(\bm{x},t;\theta)_t - \frac{\mathrm{i}\varepsilon}{2}\psi(\bm{x},t;\theta)_{\bm{x}\bm{x}} + \frac{\mathrm{i}}{\varepsilon}V(\bm{x})\psi(\bm{x},t;\theta)\right\|^2_{2,\Omega\times[0,T]},\\
		\mathcal{L}_b(\theta) &= \left\|\mathcal{P}(\psi(\bm{x},t;\theta))-\psi(\bm{x},t;\theta)\right\|^2_{2,\partial \Omega\times [0,T]}, \\
		\mathcal{L}_i(\theta) &= \left\|\psi(\bm{x},0;\theta) - \varphi(\bm{x})\exp(\mathrm{i}\phi(\bm{x})/\varepsilon)\right\|^2_{2,\Omega}. \\
	\end{split}
\end{equation*}
These loss functions measure how well $\psi(\bm{x},t;\theta)$ satisfies 
the equation (\ref{semi_sch}), the periodic boundary conditions, and 
the initial condition, respectively. \par     
\begin{figure}[H]
	\centering
	\includegraphics[width=0.7\linewidth]{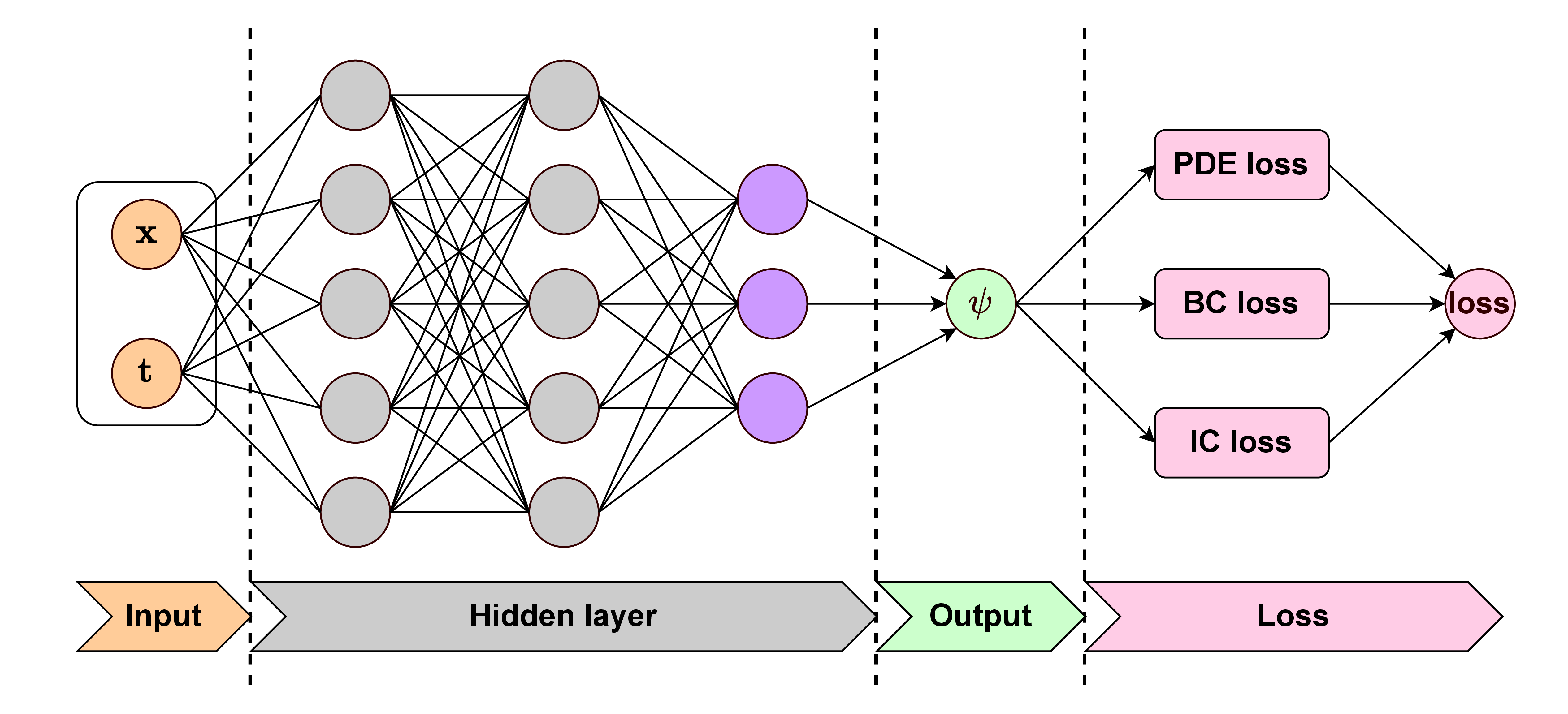}
	\caption{The structure of PINNs. Here PDE loss, BC loss, and 
	IC loss are $\mathcal{L}_r(\theta)$, $\mathcal{L}_b(\theta)$,
	and $\mathcal{L}_i(\theta)$ in equation (\ref{loss_sch}), respectively.}
	\label{pinn}
\end{figure}

The networks structure of the general PINNs is presented in 
Figure \ref{pinn}. In practice, the loss function $\mathcal{L}(\theta)$
needs to be discretized by providing a training dataset, which 
consists of  sample points $\{(\bm{x}_r^n, t_r^n)\}_{n=1}^{N_r}$, 
$\{(\bm{x}_b^n, t_b^n)\}_{n=1}^{N_b}$, 
 $ \{(\bm{x}_i^n, 0)\}_{i=1}^{N_i}$
from $\Omega\times [0, T]$, $\partial \Omega\times [0, T]$, and $\Omega \times 0$. 
We then consider the discrete loss function 
\begin{equation}
	\label{dis_loss_sch}
	\hat{\mathcal{L}}(\theta)= w_r \cdot \hat{\mathcal{L}}_r(\theta) + w_b \cdot \hat{\mathcal{L}}_b(\theta)+ w_i \cdot \hat{\mathcal{L}}_i(\theta),
\end{equation}
where
\begin{equation*}
	\begin{split}
		\hat{\mathcal{L}}_r(\theta) &= \sum_{n=1}^{N_r}\left|\bm{\psi}_{t}(\bm{x}_r^n, t_r^n;\theta) - \frac{\mathrm{i} \varepsilon}{2} \bm{\psi}_{xx}(\bm{x}_r^n, t_r^n;\theta) + \frac{\mathrm{i}}{\varepsilon}\bm{V}(\bm{x}_r^n) \bm{\psi}(\bm{x}_r^n, t_r^n;\theta)\right|^2,\\
		\hat{\mathcal{L}}_b(\theta) &= \sum_{n=1}^{N_b}\left|\mathcal{P}(\psi(\bm{x}_b^n, t_b^n;\theta)) - \psi(\bm{x}_b^n, t_b^n;\theta) \right|^2, \\
		\hat{\mathcal{L}}_i(\theta) &= \sum_{n=1}^{N_i}\left|\bm{\psi}(\bm{x}_i^n, 0) - \varphi(\bm{x}_i^n)\exp(\mathrm{i}\phi(\bm{x}_i^n)/\varepsilon)\right|^2. \\
	\end{split}
\end{equation*}
The discrete loss function (\ref{dis_loss_sch}) can  be minimized using 
stochastic gradient-based algorithm.\par

\begin{figure}[H]
	\centering
	\includegraphics[width=0.6\linewidth]{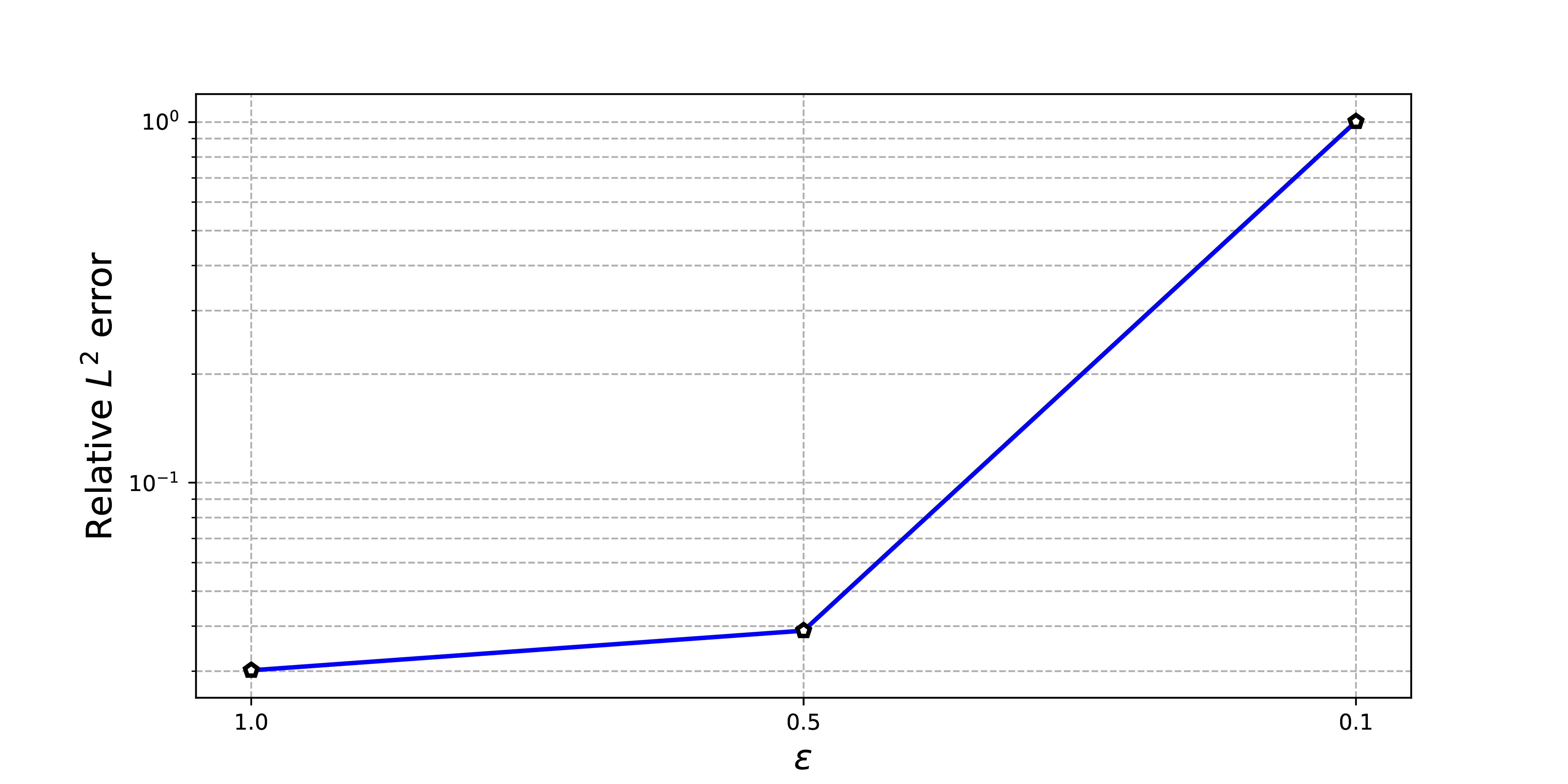}
	\caption{The Schrödinger equation (\ref{semi_sch}): the relative 
	$L^2$ errors of PINNs corresponding to different $\varepsilon$ 
	values.}
	\label{pinn1}
\end{figure}

We explore the use of PINNs to solve the one-dimensional semi-classical limit of the Schr\"{o}dinger equation (\ref{semi_sch}), 
where the initial value is given by: 
\begin{equation*}
	\psi(x,0)=\exp\biggl\{\frac{\mathrm{i}}{\varepsilon}\left[\mathrm{i} \left(x - 1\right)^2 + 2\left(x - 1\right) + \frac{1}{4}\ln \frac{2}{\pi \varepsilon}\right]\biggr\},
\end{equation*}
with $\varepsilon = $ 1.0, 0.5, or 0.1.
The computational domain is defined as $\Omega\times[0,T]=(-2\pi, 2\pi)\times[0,1]$. After training, we obtain the numerical solutions using PINNs.   
Figure \ref{pinn1} illustrates the relative $L^2$ errors of these solutions for 
$\varepsilon = $ 1.0, 0.5, and 0.1. As $\varepsilon$ decreases, the accuracy of PINNs deteriorates rapidly. 
For $\varepsilon=0.1$, the 
relative $L^2$ error exceeds 1. This outcome indicates that PINNs struggle 
to converge to accurate solutions due to the 
multi-scale effects induced by $\varepsilon$. 
To mitigate these multi-scale effects, Gaussian wave packets can be used to reformulate the original equation (\ref{semi_sch}) into an ODE system.
In the next subsection, we apply PINNs to solve this ODE system, demonstrating improved results compared to directly addressing the original multi-scale problem.

\subsection{PINNs with Gaussian wave packets}
\label{pinn_gwp}
To simplify the introduction of Gaussian wave packets, we focus on the one-dimensional case ($d=1$). This framework can be readily extended to higher-dimensional cases using tensor product grids. By expressing the solution as a sum of Gaussian wave packets, Heller \cite{heller1975time} proposed a method for efficiently solving the semi-classical limit of the Schr{\"o}dinger equation. A Gaussian wave packet solution for equation (\ref{semi_sch}) is defined as:
\begin{equation}
	\label{gwp_eq}
	\psi(x,t)=\exp\left[\frac{\mathrm{i}}{\varepsilon}\left(\alpha(t)\left(x-q(t)\right)^2+p(t)\left(x-q(t)\right)+\gamma(t)\right)\right],
\end{equation}   
where $\alpha(t)$ and $\gamma(t)$ are complex-valued functions, and $p(t)$ and $q(t)$ are real-valued functions. These parameters evolve according to the motion of a classical particle. Typically, $\gamma_{\text{re}}(0)$ is set to 0, and $\gamma_{\text{im}}(0)$ is selected to ensure the normalization of the initial wave function:
\begin{equation*}
	\int_{-\infty}^{\infty}|\psi(x,0)|^2\mathrm{d}x = 1.
\end{equation*}
For small values of $\varepsilon \ (\varepsilon \ll 1)$, 
equation (\ref{gwp_eq}) describes a highly oscillatory function 
with a Gaussian envelope centered at $q(t)$. The width of 
the envelope (i.e. its standard deviation), 
is proportional to $\sqrt{\varepsilon  /\alpha_{\text{im}}(t)}$, 
where $\alpha_{\text{im}}(t)$ denotes the imaginary part of $\alpha(t)$.
The wavelength of the oscillations at $x=q(t)$ is $2\pi\varepsilon/p(t)$. 
Additionally, the real part of $\alpha(t)$ introduces 
finer oscillations in the tails of the wave packet.\par

When the potential $V(x)$ 
is quadratic and the initial conditions is a Gaussian wave packet, Heller demonstrated that 
(\ref{gwp_eq}) serves as an exact solution of the Schrödinger 
equation (\ref{semi_sch}), provided 
the parameters satisfy the following system of ODEs:
\begin{equation}
	\label{gwp_ode}
	\begin{cases}
		\dot{q}&=p, \\
		\dot{p}&=-V'(q), \\
		\dot{\alpha}&= -2\alpha^2 - \frac{1}{2}V''(q),\\
		\dot{\gamma}&= \frac{1}{2}p^2 - V(q) + \mathrm{i} \alpha \varepsilon.
	\end{cases}
\end{equation}
For non-harmonic potentials, the theoretical analysis in \cite{heller1975time} shows that the modeling 
error of Gaussian wave packet is $O(\sqrt{\varepsilon})$.
Consequently, the Gaussian wave packet exhibits stable and accurate behavior
as $\varepsilon \to 0$.
These observations highlight Gaussian wave packets as 
a powerful tool for solving the 
semi-classical limit of the Schr{\"o}dinger equation. As reported in \cite{heller1975time, leung2009eulerian}, if the  initial data is given by the following Gaussian wave packet:
\begin{equation*}
	\label{gb_ini}
	\psi(x, 0) = \exp\left[\frac{\mathrm{i}}{\varepsilon}\left(\alpha(0)\left(x-q(0)\right)^2 + p(0)\left(x-q(0)\right) + \gamma(0)\right)\right],
\end{equation*}
the solution $\psi$ corresponds to a first-order Gaussian beam, 
as described in equation (\ref{gwp_eq}).
For general initial conditions, the solution can be approximated using a collection of Gaussian wave packets.
Inspired by Gaussian wave packets, the formulation (\ref{gwp_ode}) provides an efficient approach for solving the Schr{\"o}dinger equation in the semi-classical regime. 
However, due to the term
$\alpha(t)\left(x-q(t)\right)^2+p(t)\left(x-q(t)\right)+\gamma(t)$ in equation (\ref{gwp_eq}), which is scaled by $\frac{1}{\varepsilon}$, solving the ODE system (\ref{gwp_ode}) with a $k$-order time integration scheme results in a total truncation error of order  $\frac{(\Delta t)^k}{\varepsilon}$. This imposes an additional constraint on the time step size, potentially increasing the computational complexity. 
\par 

Next, we introduce PINNs to solve the ODE system (\ref{gwp_ode}). For $d=1$, let us denote the solutions of (\ref{gwp_ode}) as the following vector:
\begin{equation}
	\bm{y}(t) = \left(q(t), p(t), \alpha(t), \gamma(t)\right).
	\label{1dvector_gwp}
\end{equation}
Let $\bm{y}_{\theta}=\left(q_{\theta_1},\ p_{\theta_2}, \ \alpha_{\theta_3},\right.$ $\left. \gamma_{\theta_4}\right)$ 
be the outputs of PINNs, 
where $\theta = \{\theta_1, \ \theta_2, \ \theta_3, \  \theta_4\}$.
The networks of PINNs take $t\in [0,T]$ as input. 
The outputs, $q_{\theta_1}, \ p_{\theta_2}, \ \gamma_{\theta_3},$ and $\alpha_{\theta_4}$ 
are obtained by minimizing the following loss function $\hat{\mathcal{L}}(\theta)$ with stochastic gradient-based algorithm:
\begin{equation*}
		\hat{\mathcal{L}}(\theta) = w_r \cdot \hat{\mathcal{L}}_r(\theta) + w_b \cdot \hat{\mathcal{L}}_b(\theta),
\end{equation*}
where 
\begin{equation*}
	\begin{aligned}
		\hat{\mathcal{L}}_r(\theta) = &\frac{1}{N}  \sum_{i=1}^N \bigg\{ \left|\dot{q}_{\theta_1}(t_i) - p_{\theta_2}(t_i)\right|^2 + \left|\dot{p}_{\theta_2}(t_i) + V'(q_{\theta_1} (t_i))\right|^2 + \left|\dot{\alpha}_{\theta_3}(t_i) + 2 \alpha_{\theta_3}^2(t_i) + \frac{1}{2}V''(q_{\theta_1}(t_i))\right|^2  \\
		& + \left|\dot{\gamma}_{\theta_3}(t_i) - \frac{1}{2}p_{\theta_2}^2(t_i)+ V(q_{\theta_1}(t_i))-\mathrm{i}\alpha_{\theta_3}(t_i)\varepsilon\right|^2 \bigg\},
	\end{aligned}
\end{equation*}
and 
\begin{equation*}
	\hat{\mathcal{L}}_b(\theta) = \left|\bm{y}_{\theta}(0) - \bm{y}(0)\right|^2.
\end{equation*}
With the outputs of PINNs, we can construct a solution for (\ref{semi_sch}) as following
\begin{equation}
	\label{nn_gwp_semi}
	\psi(x,t;\theta)=\exp\left[\frac{\mathrm{i}}{\varepsilon}\left(\alpha_{\theta_3}(t)\left(x-q_{\theta_1}(t)\right)^2+p_{\theta_2}(t)\left(x-q_{\theta_1}(t)\right)+\gamma_{\theta_4}(t)\right)\right].
\end{equation}
According to \cite{heller1975time, leung2009eulerian}, if the total error of PINNs in solving (\ref{gwp_ode}) is ${\cal E}_t$, then the total error of solution provided by (\ref{nn_gwp_semi}) is ${\cal E}_t/\varepsilon$. 
However, the accuracy of  PINNs in solving (\ref{gwp_ode}) is generally limited.
As a result, PINNs combined with Gaussian wave packets are not sufficiently accurate for solving (\ref{semi_sch}) when the parameter $\varepsilon$ is very small. This limitation motivates the introduction of MscaleDNNs, which enhance the accuracy of the final solution in (\ref{nn_gwp_semi}).

\subsection{MscaleDNNs with Gaussian wave packets}
\label{ms_gwp}
MscaleDNNs \cite{liu2020multi} aim to reduce high frequency 
learning problems to low frequency learning problems
by using a down-scaling mapping in phase space. Let 
us consider a band-limited function $f(\bm{z})$ with $\bm{z}=(\bm{x},t)\in \mathbb{R}^{d+1}$, 
whose Fourier transform $\hat{f}(\bm{k}):={\cal F}[f(\bm{z})](\bm{k})$ has a compact support 
$\mathbb{K}\left(K_{\text{max}}\right) = \left\{\bm{k}\in \mathbb{R}^{d+1}, |\bm{k}|\leq K_{\text{max}}\right\}$. 
The compact support is then decomposed into the union of $M$ 
concentric annuli with uniform or nonuniform widths, e.g.,
\begin{equation*}
\mathbb{K}_i = \left\{\bm{k}\in \mathbb{R}^{d+1}, \ (i-1)K_0 \leq |\bm{k}|\leq i K_0\right\}, \ K_0=K_{\text{max}}/M, \ 1\leq i \leq M,
\end{equation*}
and 
\begin{equation*}
  \mathbb{K}(K_{\text{max}}) = \bigcup_{i=1}^M \mathbb{K}_i.
\end{equation*}
Based on this decomposition, $\hat{f}(\bm{k})$ can be rewritten as
\begin{equation*}
  \hat{f}(\bm{k}) = \sum_{i=1}^W \mathcal{X}_{\mathbb{K}_i}(\bm{k})\hat{f}(\bm{k})\triangleq \sum_{i=1}^W \hat{f}_i(\bm{k}),
\end{equation*}
where $\mathcal{X}_{\mathbb{K}_i}$ is the indicator function of the set $\mathbb{K}_i$ and 
$ \text{supp} \hat{f}_i(\bm{k}) \subset \mathbb{K}_i$. 
Using the inverse Fourier transform, we obtain the corresponding decomposition in the physical 
space: 
\begin{equation*}
  f(\bm{z}) = \sum_{i=1}^M f_i (\bm{z})
\end{equation*}
with 
 $  f_i(\bm{z}) = \mathcal{F}^{-1}[\hat{f}_i(\bm{k})](\bm{z}) $.

MscaleDNNs introduce $M$ sub-networks, and the $i$-th sub-network of MscaleDNNs aims to fit function $f_i(z)$ with down-scaling embedding $a_i z$ \cite{liu2020multi}. Here $a_i$ is a pre-defined parameter. The down-scaling embedding $a_i z$ converts $\hat{f}_i(\bm{k})$ into a function $\hat{f}_i^{(\text{scale})} (\bm{k}): = \hat{f}_i (a_i \bm{k})$ with a low frequency region.
The compact support of $\hat{f}_i^{(\text{scale})} (\bm{k})$ is
\begin{equation*}
  \text{supp}\hat{f}_i^{(\text{scale})}(\bm{k})\subset \left\{\bm{k} \in \mathbb{R}^{d+1}, 
  \frac{(i-1)K_0}{a_i}\leq |\bm{k}| \leq \frac{iK_0}{a_i}\right\}.
\end{equation*}
In the physical space, the down-scaling is denoted as
\begin{equation*}
  f_i^{(\text{scale})}(\bm{z}) = \frac{1}{a_i^{d+1}}  f_i \left(\frac{1}{a_i} \bm{z}\right) \quad \text{or} \quad f_i(\bm{z}) = a_i^{d+1} f_i^{(\text{scale})}(a_i \bm{z}).
\end{equation*}

In MscaleDNNs, $a_i$ is chosen to be sufficiently large 
such that $f_i^{(\text{scale})}(\bm{z})$
exhibits a low-frequency spectrum. 
Typically, the value of $a_i$ is set to $2^{i-1}$.
According to the F-Principle
in conventional DNNs \cite{xu2020frequency}, 
MscaleDNNs facilitate rapid learning for $f_i^{(\text{scale})}(\bm{z})$, 
denoted as $f_{\theta_i}$, where $\theta_i$ represents 
the parameters of the MscaleDNNs.
Finally, MscaleDNNs approximate $f(\bm{z})$ as: 
\begin{equation}
  f_\theta(\bm{z}) = \sum_{i=1}^M a_i^{d+1} f_{\theta_i}(a_i \bm{z}).
  \label{eq21_1}
\end{equation}
In summary, MscaleDNNs transform the original 
input data $\bm{z}$ into multi-scale representation
$\{a_1\bm{z}, \dots, a_M\bm{z}\}$, which is then 
fed into the first hidden layer of the neural network. 
Numerical simulations 
reported in \cite{liu2020multi} 
validate that MscaleDNNs provide an efficient, mesh-free, 
and straightforward approach for solving multi-scale PDEs. Furthermore, the numerical experiments in section \ref{compare_pinn_ms} will demonstrate the significant accuracy advantages of MscaleDNNs over conventional DNNs in solving the ODE system (\ref{gwp_ode}).

\section{DeepONets for the semi-classical limit of the Schr{\"o}dinger equation}\label{sec3}
As discussed in the previous section, DNNs have demonstrated excellent performance in solving the Schr{\"o}dinger equation with fixed initial conditions.  
However, in many applications, such as those explored in \cite{russo2013gaussian} and \cite{faou2009computing}, there is a need to address the semi-classical limit of the Schr{\"o}dinger equation with varying initial conditions. This presents a crucial challenge: the ability to simultaneously and efficiently solve the Schr\"{o}dinger equation (\ref{semi_sch}) for multiple initial conditions. When initial conditions change, DNNs typically require retraining to achieve highly accurate solutions, which significantly hampers their performance in handling multiple initial conditions efficiently. To overcome this limitation, we introduce DeepONets \cite{2021LuLu} in this section as an alternative approach for solving the Schr\"{o}dinger equation (\ref{semi_sch}) in 
subsection \ref{pd}, and discuss how to integrate them with Gaussian wave packets in subsection \ref{pd_gwp}.

\subsection{Physics-Informed Deep Operator Networks}
\label{pd}
\begin{figure}[htbp]
	\centering
	\includegraphics[width=0.7\linewidth]{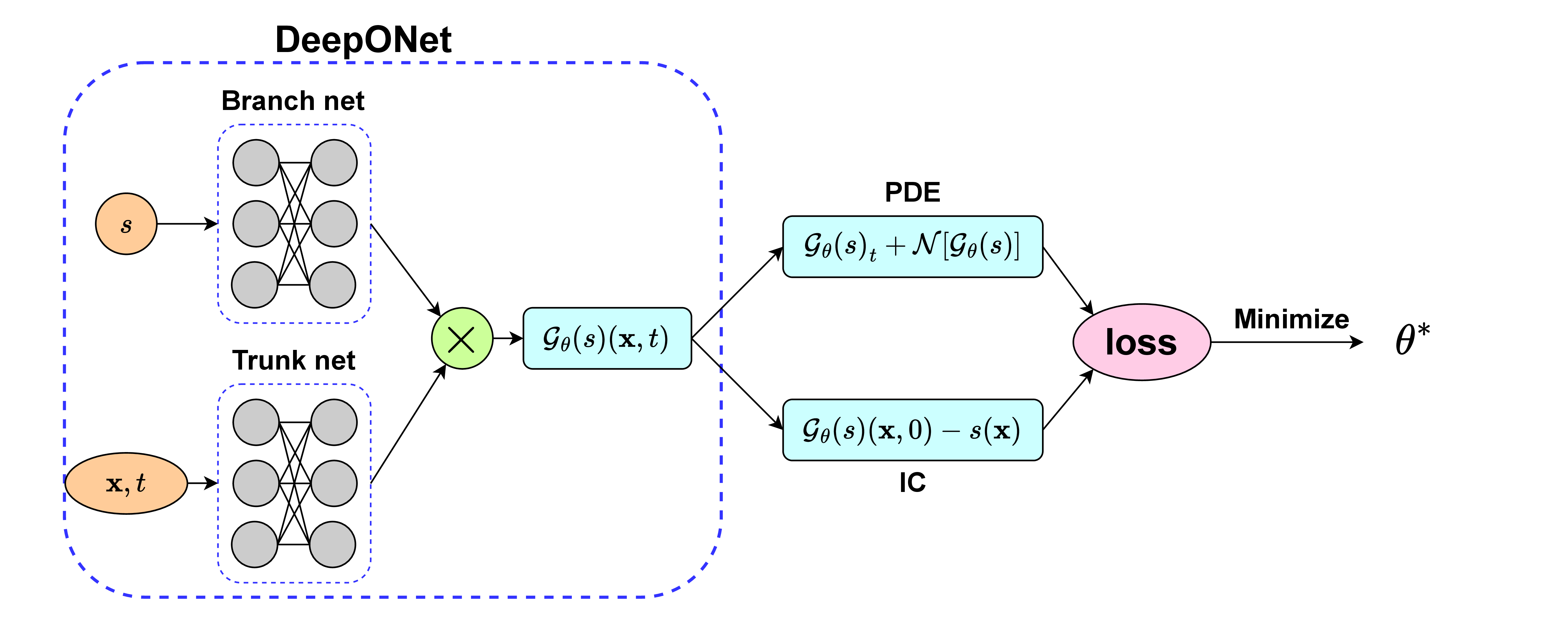}
	\caption{Physics-Informed DeepONets: The DeepONets architecture \cite{2021LuLu} consists of two sub-networks: the branch network and the trunk network. The outputs of these sub-networks are combined using a dot product to produce a continuously differentiable representation of the output function. The optimal parameters are obtained by minimizing a loss function defined by the governing PDEs. }
	\label{fig:deeponet}
\end{figure}

DeepONets are specialized deep learning architectures designed to learn abstract, nonlinear operators between 
infinite-dimensional function spaces. 
In the context of time-dependent problems, DeepONets excel at mapping initial conditions to solutions over a given time interval.
Let us begin with a general time-dependent PDEs of the following form: 
\begin{equation}
	\label{gen_td_pde}
	\left\{ 
		\begin{aligned}
			&\psi_t + \mathcal{N}[\psi]=0, \quad (\bm{x},\,t)\in \Omega\times (0, T], \\
			&\psi(\bm{x}, 0) = s(\bm{x}), \quad   \bm{x}\in \Omega,\\
		\end{aligned}
	\right.
\end{equation}
with periodic boundary conditions prescribed on the boundary of $\Omega$. 
The spatial domain is denoted by $\Omega$, 
and $T$ represents the time horizon.
Assuming that for any $s(\bm{x})\in \mathcal{S}$, 
there exists a unique solution 
$\psi(\bm{x},t)\in \bm{\Psi}$ for problem (\ref{gen_td_pde}). Here $\mathcal{S}$ and $\bm{\Psi}$ are infinite-dimensional function spaces. The PDE solution operator  
$\mathcal{G}:\mathcal{S}\to \bm{\Psi}$ for (\ref{gen_td_pde}) is defined as:
\begin{equation*}
	\label{op}
	\mathcal{G}[s] = \psi.
\end{equation*}
DeepONets are employed to approximate this solution operator as $\mathcal{G}_{\theta}$, 
where $\theta$ represents all trainable parameters.
As shown in Figure \ref{fig:deeponet}, 
DeepONets consist of two key components:
the branch networks and trunk networks.
For fixed spatial locations $\{\bm{x}_i\}_{i=1}^m$, the branch networks take the values of initial function $s(\bm{x}_i)$
as input and output a feature embedding 
$[b_1, b_2, \dots, b_J]^T \in \mathbb{R}^J$. 
The trunk networks accept continuous coordinates 
$\bm{x}$ and $t$ as inputs and produce a feature embedding 
$[c_1, c_2,\dots,c_J]^T \in \mathbb{R}^J$.
The final output of DeepONets is obtained by combining 
these embeddings through an inner product:
\begin{equation*}
	\mathcal{G}_{\theta}[s](\bm{x}, t) = \sum_{j=1}^J \underbrace{b_j(s(\bm{x}_1), s(\bm{x}_2), \dots, s(\bm{x}_m))}_{\text{branch}}\underbrace{c_j(\bm{x}, t)}_{\text{trunk}}.
\end{equation*}
If the solution of (\ref{gen_td_pde}) is a 
vector-valued function, 
a natural extension of the above output is:
\begin{equation}
	\label{multi_deeponet}
	\mathcal{G}_{\theta}=(\mathcal{G}_{\theta}^{(1)}, \dots,  \mathcal{G}_{\theta}^{(I)}),
\end{equation}
where 
\begin{equation*}
	\mathcal{G}_{\theta}^{(i)}[s](\bm{x}, t) = \sum_{j=J_{i-1}+1}^{J_i} \underbrace{b_j(s(\bm{x}_1), s(\bm{x}_2), \dots, s(\bm{x}_m))}_{\text{branch}}\underbrace{c_j(\bm{x}, t)}_{\text{trunk}}, \quad i=1,\dots,I,
\end{equation*}
and $0=J_0<J_1<\cdots <J_I=J$. 
The resulting output is a vector-valued 
function with $I$ components. \par

The optimal parameters $\theta^*$ of the DeepONets is usually obtained by minimizing a loss function defined by the governing PDEs. When the solution values to equation (\ref{gen_td_pde}) at several locations are available, for a given initial condition $s(\bm{x})$, the loss function is defined as:
\begin{equation}
	\label{eq11}
	\hat{\mathcal{L}}(s, \theta)=\frac{1}{P}\sum_{j=1}^P\left|\mathcal{G}_{\theta}[s](\bm{x}_j, t_j)-\psi(\bm{x}_j, t_j)\right|^2,
\end{equation}
where $\{\psi(\bm{x}_j, t_j)\}_{j=1}^P$ denotes 
the solution to equation (\ref{gen_td_pde}) with the initial function $s(\bm{x},t)$, evaluated at $\{\bm{x}_j, t_j\}_{j=1}^P\in\Omega\times[0,T]$.
The loss function for the DeepONets is given by:
\begin{equation*}
	\hat{\mathcal{L}}(\theta) = \frac{1}{N}\sum_{i=1}^N \hat{\mathcal{L}}(s^{(i)}, \theta),
\end{equation*} 
where $\{s^{(1)}, s^{(2)}, \dots, s^{(N)}\}$ are 
randomly sampled from $\mathcal{S}$. \par

The loss function described above relies solely on
data and cannot be directly applied to solve a set of PDEs without internal observations.
To overcome this limitation, Wang \textit{et al.} \cite{2021wang} 
proposed Physics-Informed DeepONets, a framework analogous to PINNs.
This approach leverages automatic differentiation to enforce the governing PDEs as constraints on the outputs of the DeepONets model.
Specifically, the loss function for Physics-Informed DeepONets can be defined as:
\begin{equation*}
	\hat{\mathcal{L}}(\theta) = \hat{\mathcal{L}}_{\text{operator}}(\theta) + \hat{\mathcal{L}}_{\text{physics}}(\theta),
\end{equation*}
where $\mathcal{L}_{\text{operator}}(\theta)$ enforces boundary and initial conditions, 
and $\mathcal{L}_{\text{physics}}(\theta)$ ensures the 
satisfaction of the PDE's internal constraints.
For simplicity, we focus on the loss function associated with the initial conditions, noting that boundary conditions can be handled in a similar manner.
The operator loss function can be defined as: 
\begin{equation*}
	\hat{\mathcal{L}}_{\text{operator}}(\theta) = \frac{1}{N}\sum_{i=1}^N \hat{\mathcal{L}}(s^{(i)}, \theta) = \frac{1}{NP} \sum_{i=1}^N\sum_{j=1}^P \left|\mathcal{G}_{\theta}[s^{(i)}](\bm{x}^{(i)}_j, 0) - s(\bm{x}^{(i)}_j)\right|^2,
\end{equation*}
where $s^{(i)}$  is 
randomly sample from $\mathcal{S}$, and 
$(\bm{x}^{(i)}_j,0)$ denotes the 
sampling point for the initial condition.
For a given initial function $s^{(i)}$, 
we can define:
\begin{equation*}
	\hat{\mathcal{L}}_{\text{physics}}(\theta)[s^{(i)}] = \frac{1}{Q} \sum_{j=1}^Q \left|\left[\mathcal{G}_{\theta}[s^{(i)}](\bm{x}_{r,j}^{(i)},t_{r,j}^{(i)})\right]_t + \mathcal{N}\left[\mathcal{G}_\theta[s^{(i)}](\bm{x}_{r,j}^{(i)},t_{r,j}^{(i)})\right]\right|^2,
\end{equation*}
where $(x_{r,j}^{(i)}, t_{r,j}^{(i)}) \in \Omega \times [0, T]$. The total physics loss function is then given by:
\begin{equation*}
	\begin{aligned}
		\hat{\mathcal{L}}_{\text{physics}}(\theta) &=\frac 1 {N}\sum_{i=1}^{N} \hat{\mathcal{L}}_{\text{physics}}(\theta)(s^{(i)}) \\
		& = \frac{1}{NQ}\sum_{i=1}^N\sum_{j=1}^Q \left|\left[\mathcal{G}_{\theta}[s^{(i)}](\bm{x}_{r,j}^{(i)},t_{r,j}^{(i)})\right]_t + \mathcal{N}_x\left[\mathcal{G}_\theta[s^{(i)}](\bm{x}_{r,j}^{(i)},t_{r,j}^{(i)})\right]\right|^2.
	\end{aligned}
\end{equation*}

As demonstrated in \cite{2021wang, 2023wanglong}, 
physics-informed DeepONets are 
capable of learning the solution operator for parametric PDEs 
in an entirely self-supervised manner, without 
the need for any paired input-output observations. Figure \ref{fig:deeponet} illustrates
the fundamental framework of physics-informed DeepONets.  
However, the presence of a small parameter, $\varepsilon$, 
in the semi-classical limit of the Schr{\"o}dinger equation 
introduces significant computational challenges.
To address these challenges, we propose integrating Gaussian wave
packets with physics-informed DeepONets to develop
a more accurate solution operator, as discussed in the next subsection.

\subsection{Physical-Informed DeepOnets with Gaussian wave packets}
\label{pd_gwp}

Let us consider the Schr{\"o}dinger equation (\ref{semi_sch}) with the following Gaussian initial conditions:
\begin{equation*}
	\psi(x, 0) = \exp\left[\frac{\mathrm{i}}{\varepsilon}\left(\alpha(0)\left(x-q(0)\right)^2 + p(0)\left(x-q(0)\right) + \gamma(0)\right)\right],
\end{equation*}
where $q(0), p(0), \alpha(0),$ and $ \gamma(0)$ 
are parameters that may take on varying values.            
Our goal is to identify an operator 
\begin{equation*}
	\mathcal{G}: \psi(\bm{x},0) \to \psi(\bm{x},t),
\end{equation*}
that maps these varying initial conditions to corresponding solutions  at later times. 
As mentioned in subsection \ref{pinn_gwp}, when the 
equation (\ref{semi_sch}) has the following initial value (consider 1D case for convenience):
\begin{equation}
	\label{ob1}
	\psi(x,0)=\exp\left[\frac{\mathrm{i}}{\varepsilon}\left(\alpha(0) \left(x - q(0)\right)^2 + p(0)\left(x - q(0)\right) + \gamma(0)\right)\right],
\end{equation}
its solution can be represented as:
\begin{equation}
	\label{ob2}
	\psi(x,t) = \exp\left[\frac{\mathrm{i}}{\varepsilon}\left(\alpha(t) \left(x - q(t)\right)^2 + p(t)\left(x - q(t)\right) + \gamma(t)\right)\right].
\end{equation}
Consequently, the solution operator $\mathcal{G}$ can be decomposed as follows:
\begin{equation*}
	\mathcal{G}: \psi(x,0) \overset{\eqref{ob1}}{\longrightarrow} \bm{y}(0) \overset{\mathcal{G}'}{\longrightarrow} \bm{y}(t) \overset{\eqref{ob2}}{\longrightarrow}  \psi(x,t),
\end{equation*}
where $\bm{y}(t) = \left(q(t), p(t), \alpha(t), \gamma(t)\right)$ and the operator $\mathcal{G}'$ is defined by:
\begin{equation*}
	\mathcal{G}': \bm{y}(0) \to \bm{y}(t),
\end{equation*}

\begin{figure}[htbp]
	\centering
	\includegraphics[width=0.7\linewidth]{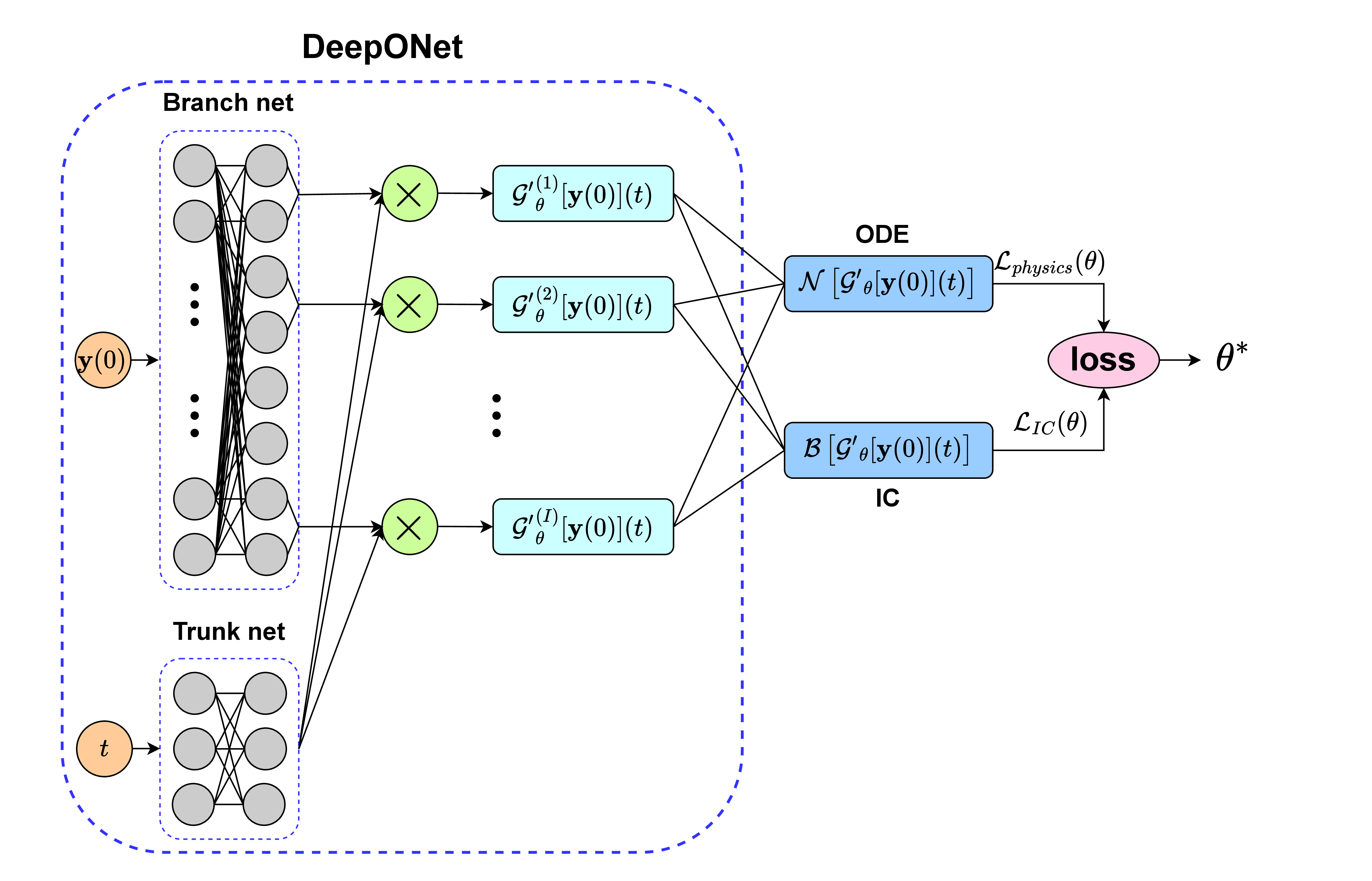}
	\caption{Physics-informed DeepONets structure of $\mathcal{G}'_{\theta}$.
    The branch networks extract the initial values 
	of $\bm{y}(0)$, while the trunk networks 
	only take the input coordinate $t$ at which 
	the output functions are evaluated. }
	\label{fig:phydeeponet}
\end{figure}

Thus, to reduce the computational cost and improve the accuracy of physics-informed DeepONets, we learn an approximate operator $\mathcal{G}'_{\theta}$ instead of solution operator $\mathcal{G}_{\theta}$ and then reconstruct the solution 
using the output of $\mathcal{G}'_{\theta}$ and \eqref{ob2}.
Compared to $\mathcal{G}_{\theta}$, the inputs for the branch and trunk networks of the
physics-informed DeepONets $\mathcal{G}'_{\theta}$
are $\bm{y}(0)$ and $t$, which are independent of $\bm{x}$. This reduces the dimensionality of the inputs and enhances the performance of physics-informed DeepONets. Similar to \eqref{multi_deeponet}, we define $\mathcal{G}'_{\theta}$ as:
\begin{equation}
	\label{eqd}
	\mathcal{G'}_\theta = \left(\mathcal{G'}_{\theta}^{(1)}, \dots, \mathcal{G'}_{\theta}^{(I)}\right),
\end{equation}
where 
\begin{equation}
	\mathcal{G'}_{\theta}^{(i)}\left[\bm{y}(0)\right](t) = \sum_{j=J_{i-1}+1}^{J_i} \underbrace{b_j[\bm{y}(0)]}_{\text{branch}}\underbrace{c_j(t)}_{\text{trunk}}, \quad i=1,\dots,I.
\end{equation}
The loss function for $\mathcal{G}'_{\theta}$ is then defined as:
\begin{equation*}
	\hat{\mathcal{L}}(\theta) = \hat{\mathcal{L}}_{\text{boundary}}(\theta) + \hat{\mathcal{L}}_{\text{physics}}(\theta),
\end{equation*}
where 
\begin{equation*}
	\label{phy_deeponet_loss}
	\hat{\mathcal{L}}_{\text{boundary}}(\theta) = \frac{1}{N} \sum_{i=1}^N \sum_{j=1}^I \left|\mathcal{G'}_{\theta}^{(j)}\left[\bm{y}^{(i)}(0)\right](0) - y^{(i)}_j(0)\right|^2,
\end{equation*}
and 
\begin{equation}
		\hat{\mathcal{L}}_{\text{physics}}(\theta) = \frac{1}{NQ}  \sum_{i=1}^N \sum_{j=1}^Q  \left\{\left|\mathcal{N}^{(1)}\left(\mathcal{G'}_{\theta}\left[\bm{y}^{(i)}(0)\right](t_j^{(i)})\right)\right|^2 + \cdots + \left|\mathcal{N}^{(I)}\left(\mathcal{G'}_{\theta}\left[\bm{y}^{(i)}(0)\right](t_j^{(i)})\right)\right|^2  \right\}.
\end{equation}
Here, $(\mathcal{N}^{(1)}, \dots, \mathcal{N}^{(I)})$ represent the operators in the ODE 
system (\ref{gwp_ode}) and $t_j^{(i)} \in [0, T]$. The structure of the physics-informed DeepONets for $\mathcal{G}'_{\theta}$ is displayed in Figure~\ref{fig:phydeeponet}. 

\begin{remark}
Due to the limitations of a single Gaussian wave packet, its accuracy 
is often restricted by the Ehrenfest time, beyond which the model error 
can grow significantly, especially in the case of non-harmonic potentials. 
As a result, the neural network solvers proposed in this paper are not 
suitable for long-time simulations, even though they can accurately solve 
the associated ODEs over extended time intervals. Performing long-time 
simulations of the Schr{\"o}dinger equation near the semi-classical 
limit remains a challenging task. One promising direction is to 
incorporate neural networks that dynamically learn corrections to 
the Gaussian wave packets using either data or physical insights. 
Furthermore, leveraging error estimation to embed adaptive strategies 
into the network architecture could enhance both the efficiency and 
accuracy of the neural network approaches.
\end{remark}

\section{Numerical experiments for PINNs and MscaleDNNs with Gaussian wave packets}
\label{num_exp}
In this section, we evaluate the performance of  PINNs with Gaussian wave packets and MscaleDNNs with Gaussian wave packets in solving the semi-classical limit of the Schr{\"o}dinger equation. We consider a range of examples,
including one-dimensional (1D), two-dimensional (2D), 
and four-dimensional (4D) cases \cite{russo2013gaussian, russo2014gaussian, bao2002time, faou2009computing}. Periodic boundary conditions are applied across all test cases. The sample points are selected to be sufficiently dense to capture the highly oscillatory solutions.
To assess precision, we need a reference solution $\psi_{ref}$. In the absence of an exact solution, we use a fourth-order scheme \cite{chin2001fourth} for solving the Schrödinger equation with non-harmonic potential and the fourth-order Runge--Kutta method with Gaussian wave packets for solving the Schrödinger equation with the harmonic potential to obtain $\psi_{ref}$. For 
error evaluation, we primarily utilize 
the relative $L^2$ error, defined as:
\begin{equation*}
    E_{\text{rel}}(\psi_{\text{pred}},\psi_{\text{ref}}) = \frac{\|\psi_{\text{pred}} - \psi_{ref} \|_{L^2(\Omega\times [0,T])}}{\| \psi_{ref} \|_{L^2(\Omega\times [0,T])}} = \frac{\sqrt{\int_{\Omega\times [0, T]} |\psi_{\text{pred}}(\bm{x},t) - \psi_{\text{ref}}(\bm{x},t)|^2 \, \mathrm{d}\bm{x}\mathrm{d}t}}{\sqrt{\int_{\Omega\times[0,T]} |\psi_{\text{ref}}(\bm{x},t)|^2 \, \mathrm{d}\bm{x}\mathrm{d}t}}.
\end{equation*}
In this section, we primarily focus on the performance of PINNs 
and MscaleDNNs with Gaussian wave packets. An extension to more 
general initial conditions using Hagedorn wave packets is provided 
in \ref{ha_wave_packets}.

\subsection{Comparison of PINNs and MscaleDNNs in solving ODEs system}
\label{compare_pinn_ms}
Let us use MscaleDNNs to solve (\ref{gwp_ode}) and compare their performance with that of PINNs.
In (\ref{gwp_ode}), we set the potential $V(x)=0$ and $\epsilon=0.01$.
The initial values are set as: $q(0)=1,\ p(0)=2, \ \alpha(0)=\mathrm{i},$ and $  \gamma(0)=-\frac{1}{4}\log(\frac{200}{\pi})\mathrm{i}$. 
With these setting, the exact solution for (\ref{gwp_ode}) is given by: 
\begin{equation*}
	\begin{cases}
		q(t)&=1+2t, \\
		p(t)&=2, \\
		\alpha(t)&= \frac{2t+\mathrm{i}}{4t^2+1},\\
		\gamma(t)&= 2t-0.005\arctan(2t)+\frac{1}{4}(\log(4t^2+1)-\log(\frac{200}{\pi}))\mathrm{i}.
	\end{cases}
\end{equation*} 
The 
architecture of the PINNs model is specified by [1, 100, 400, 400, 400, 400, 6], 
denoting the number of neurons in the input layer, 
four hidden layers, and the output layer, respectively.
In MscaleDNNs, we set scaling coefficients as 
\begin{equation}
	\label{scale_coe}
	\{a_1,\ a_2,\ \ldots,\ a_{100}\}=\{0.1,\ 0.2,\ \ldots,9.9,\ 10.0\},
\end{equation}
and the network architecture is 
defined by [100, 400, 400, 400, 400, 6], 
representing the number of neurons in each 
layer from input to output.
The relative $L^2$ errors are presented in Table \ref{ode_table}, 
indicating that the accuracy of MscaleDNNs surpasses that of 
PINNs by more than two orders of magnitude. Based on the 
Gaussian wave packets, the final numerical $L^2$ error in 
solving (\ref{semi_sch}) is scaled by a factor of 
$\frac 1 \varepsilon$ relative to the error of the ODE 
system. Therefore, even though the relative $L^2$ errors 
of PINNs range from 1e-05 to 1e-03, they are still insufficiently accurate for solving (\ref{semi_sch}), particularly for small values of $\varepsilon$. 
In the next subsection, we provide a direct comparison of 
PINNs and MscaleDNNs with Gaussian wave packets in solving (\ref{semi_sch}).

\begin{table}[H]
	\centering
    \caption{The relative $L^2$ error of $q(t), \ p(t), \ \alpha(t), \ \gamma(t)$ for PINNs and MscaleDNNs.}
	\tabcolsep=0.40cm
	\begin{tabular}{ccccc}
		\hline
		different part of ODE system& $q(t)$ & $p(t)$ & $\alpha(t)$ & $\gamma(t)$\\
		\hline
		Relative $L^2$ error of PINNs& 3.287e-05 & 2.585e-04 & 3.105e-03 & 2.864e-04\\
		\hline
		Relative $L^2$ error of MscaleDNNs&2.928e-07 & 4.141e-06 & 1.941e-05 & 4.161e-06\\
		\hline
	\end{tabular}
	\label{ode_table}
\end{table}

\begin{figure}[H]
	\centering
	\subfloat[]{\includegraphics[width=0.4\linewidth]{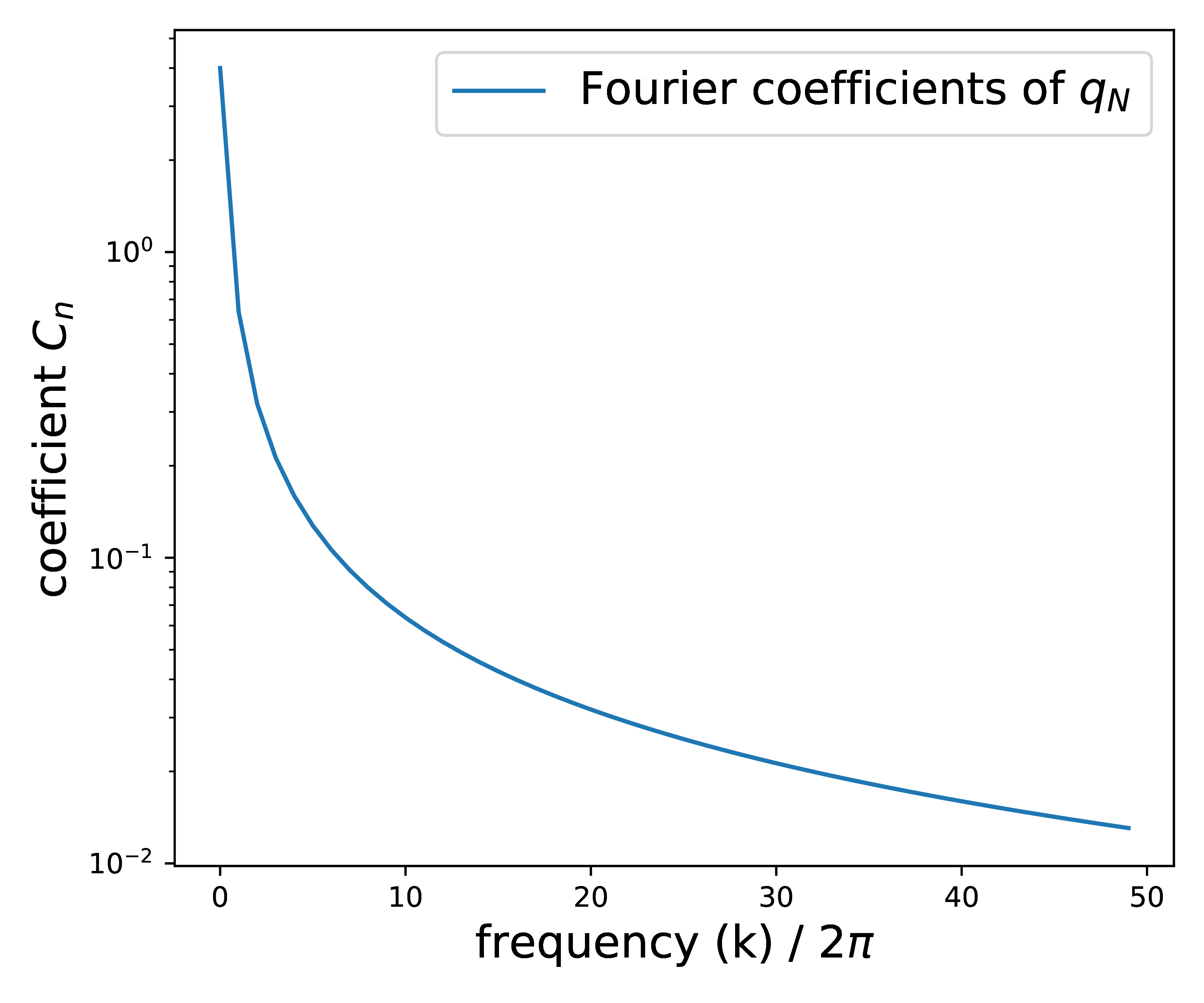}}\quad
	\subfloat[]{\includegraphics[width=0.4\linewidth]{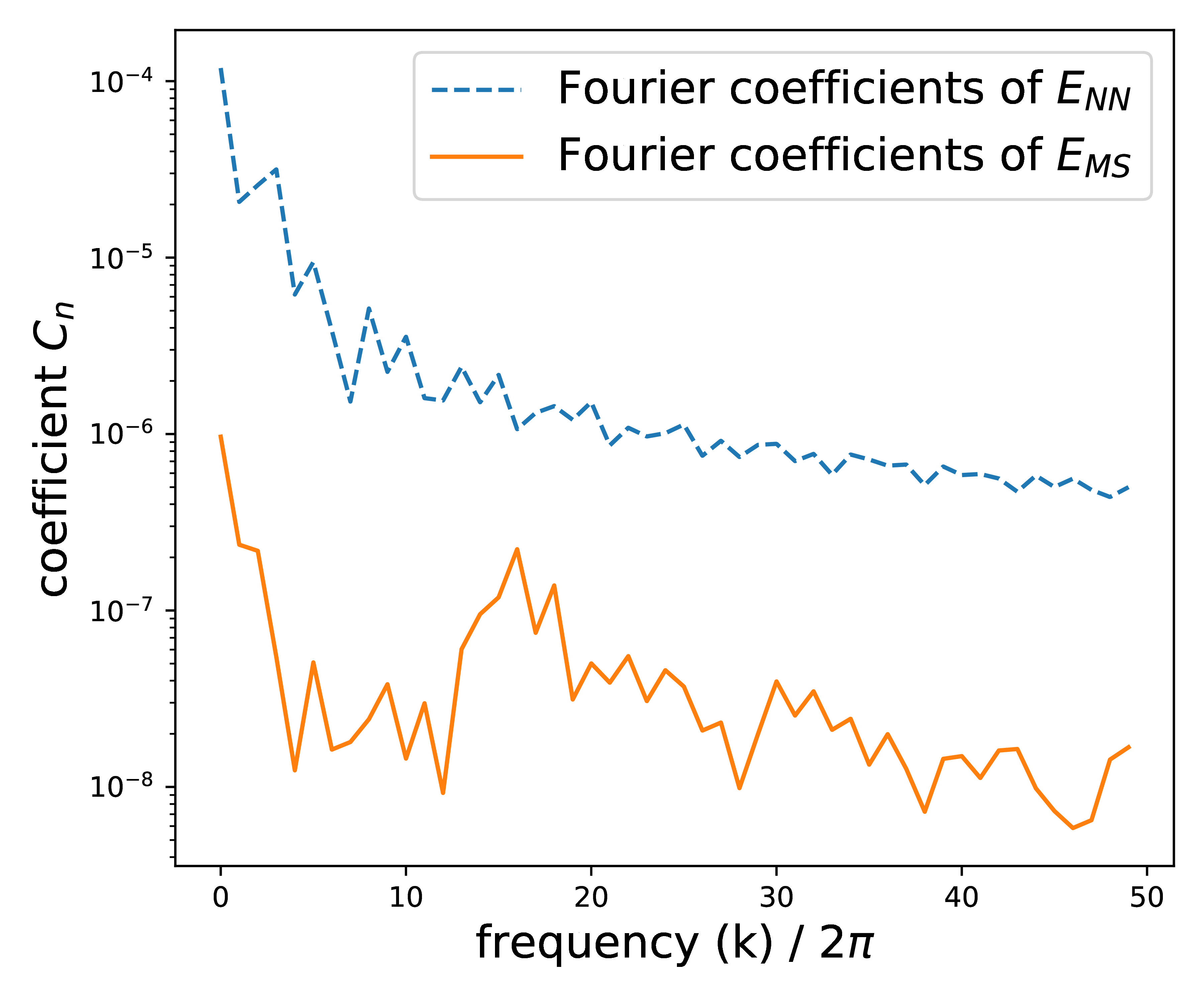}}\hfill
	\caption{(a) Distribution of Fourier coefficients $C_n$ for the exact solution $q(t)$.
	(b) The blue dotted line and the orange solid line represent the distribution of absolute errors in the Fourier coefficients obtained using PINNs and MscaleDNNs, respectively.}
	\label{fre_q}
\end{figure}

Let us define the absolute error of $q(t)$ for PINNs and MscaleDNNs as $E_{\text{NN}}=|q_{\text{NN}}(t) - q(t)|$ and $E_{\text{MS}}=|q_{\text{MS}}(t) - q(t)|$, respectively.
The discrete Fourier transform on $q(t)$ is given as follows:
\begin{equation*}
	q_{N}(t) = \sum_{n=-N}^{N} C_n e^{2\pi \mathrm{i}n t},
\end{equation*}
where $N=100$. The Fourier coefficients of the exact solution $q(t)$ are reported in Figure \ref{fre_q}(a), exhibiting a long-tail distribution, a common characteristic for many functions.
The corresponding Fourier coefficients for $E_{\text{NN}}$ and $E_{\text{MS}}$ are presented in \ref{fre_q}(b).
According to Figure \ref{fre_q}(a), the primary information of $q(t)$ is concentrated in the low-frequency band, while a small portion of the frequency content resides in the high-frequency 
region. The dominance of low-frequency components explains why PINNs can achieve a reasonable accurate solution with relative $L^2$ error 3.287e-05.

The distribution of Fourier coefficients of $E_{\text{NN}}$ follows a similar trend to that of $q(t)$. However, after incorporating down-scaling embeddings, the MscaleDNNs effectively shift the high frequencies to low frequencies, resulting in a more uniform distributed set of Fourier coefficients for $E_{\text{MS}}$.   
This transformation enables MscaleDNNs to achieve an improvement in accuracy by two orders of magnitude compared to PINNs.

\begin{figure}[H]
	\centering
	\includegraphics[width=0.6\linewidth]{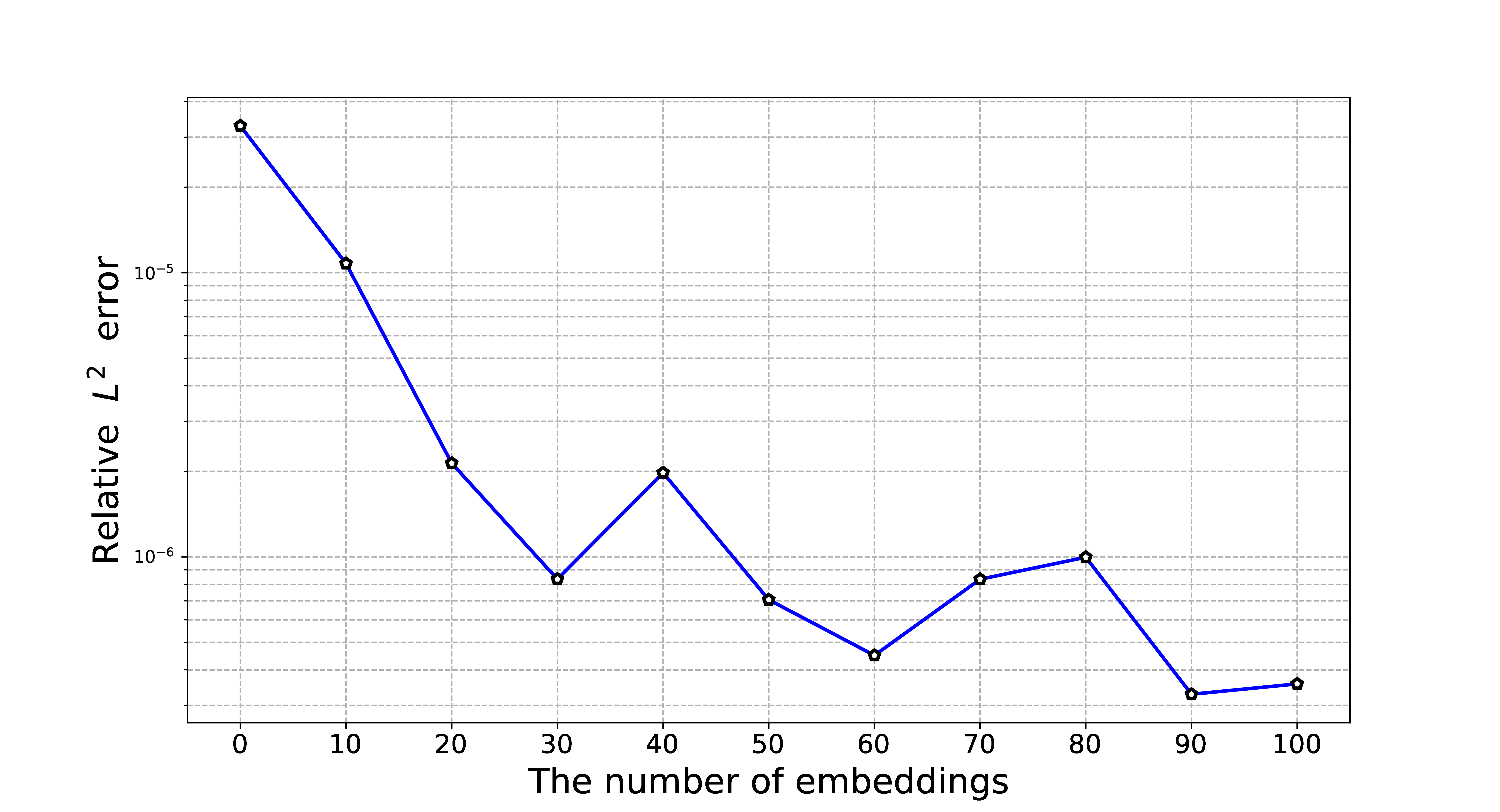}
	\caption{The relative $L^2$ errors of $q(t)$ computed using MscaleDNNs with varying numbers of embeddings. Here, $0$ corresponds to the results obtained using PINNs.}
	\label{scale_error}
\end{figure}

To investigate the impact of the number of embeddings on the performance of MscaleDNNs, we perform experiments with the number of embeddings varying from 10 to 100 in increments of 10.
The relative $L^2$ errors for these ten simulations are presented in Figure \ref{scale_error}.  From these results, we observe that
increasing the number of embeddings generally enhances the accuracy of MscaleDNNs. Due to the long-tail distribution of the Fourier coefficients of $q(t)$, MscaleDNNs achieve optimal performance when the number of embeddings approaches 100. Consequently, in the remainder of this paper, we consistently employ 100 embeddings in MscaleDNNs to ensure highly accurate solutions.

\subsection{1D examples}
\label{1d_example}
In the following two 1D examples, we consider the domain $\Omega\times[0,T]= (-\pi,\pi)\times[0,1]$.
For PINNs with Gaussian wave packets, the input consists of a single variable, $t$, while the output includes multiple components corresponding to $q$, $p$, $\alpha$, and $\gamma$.  
Note that the output dimension is not four because
$\alpha$ and $\gamma$ are complex functions that need to be separated into real and imaginary parts. As a result, 
the neural network has a layer configuration of [1, 100, 400, 400, 400, 6].
In contrast, for MscaleDNNs with Gaussian wave packets, the input consists of multiple embeddings $[0.1t, 0.2t, \dots, 10t]$ rather than the raw time variable $t$.
Consequently, the network architecture is [100, 400, 400, 400, 6], specifying the number of neurons in each layer.
Aside from this distinction, all other settings remain consistent.
\par

The network initialization 
follows the Glorot normal scheme 
\cite{glorot2010understanding}. All networks are 
trained using the Adam optimizer \cite{kingma2014adam} 
with default settings. We employ an initial learning rate of $0.001$ 
and train for 100,000 epochs.
In all simulations, the activation function is 
the Softened Fourier Mapping activation function, 
defined as $0.5\sin(x) + 0.5\cos(x)$ \cite{li2023subspace}.
To reduce the training cost and enhance stability, we set 
the training batch size to 1000.

\subsubsection{Harmonic potential: $V(x) = \frac{1}{2}x^2$.}
\label{1d_har}
Consider the harmonic potential $V(x)=\frac{1}{2}x^2$ and the following initial condition:
\begin{equation*}
	\psi(x, 0) = \exp\left[\frac{\mathrm{i}}{\varepsilon}\left(\alpha(0)\left(x - q(0)\right)^2 + p(0)\left(x - q(0)\right) + \gamma(0)\right)\right],
\end{equation*}
where $q(0)=1, \ p(0) = 2, \ \alpha(0) = 0.5i, \ \gamma_{\text{re}}(0) =0$ 
and $\gamma_{\text{im}}(0)$ is chosen
such that $\int_{-\infty}^{\infty} |\psi(x, 0)|\mathrm{d}x=1$.
We evaluate the performance of PINNs and MscaleDNNs 
with Gaussian wave packets for various values of $\varepsilon$, 
specifically $\varepsilon = \frac{4}{25}, \ \frac{1}{25}, \ \frac{1}{100}, \ \frac{1}{400}, \ \frac{1}{1600}$, and $ \frac{1}{6400}$.
To evaluate the accuracy of MscaleDNNs with respect to the number of sample points, we consider the harmonic potential $V(x)=\frac 1 2 x^2$ as a test case and conduct simulations with varying numbers of sample points over the time interval [0, 1]. The relative $L^2$ errors from these simulations are presented in Table \ref{NoSPtab2}. The results indicate that 6,400 sample points are sufficient to achieve accurate solutions for all cases with $\varepsilon\leq \frac 1{6400}$. Therefore, to eliminate the influence of sampling density on solution accuracy, we use 10,000 sample points over the interval 
[0,1] in all subsequent experiments.
The results of the relative $L^2$ errors are summarized in Table \ref{table1}. 
The most striking observation is the 
significant performance advantage 
of MscaleDNNs over PINNs. 
MscaleDNNs consistently achieve improvements exceeding two orders of magnitude, highlighting the effectiveness of integrating MscaleDNNs with Gaussian wave packets.
This demonstrates the superior capability of MscaleDNNs in capturing the intricate dynamics of the Schrödinger equation, particularly for small values of 
$\varepsilon$, where the solutions exhibit highly oscillatory behavior.

For MscaleDNNs with Gaussian wave packets, the relative $L^2$ errors exhibit a linear increase as $\varepsilon$ decreases. In the case of harmonic potentials and the specified initial condition, as detailed in subsection \ref{pinn_gwp},
Gaussian wave packets introduce zero model error. As a result, 
the primary source of relative $L^2$ error stems from solving the ODE 
system (\ref{gwp_ode}) using MscaleDNNs. Additionally, to derive the final solution, the solution of this ODE system is integrated into the Gaussian wave packet solution defined by (\ref{gwp_eq}), which includes the term $\frac{\mathrm{i}}{\varepsilon}(\alpha (x - q)^2 + p(x - q) + \gamma)$. This formulation introduces a factor of $\frac{1}{\varepsilon}$ into the final relative $L^2$ errors. Therefore, we can conclude that MscaleDNNs achieve consistent errors for ODE system (\ref{gwp_ode}), which remain unaffected by variations in $\varepsilon$. 

\begin{figure}[htbp]
	\centering
	\includegraphics[width=0.6\linewidth]{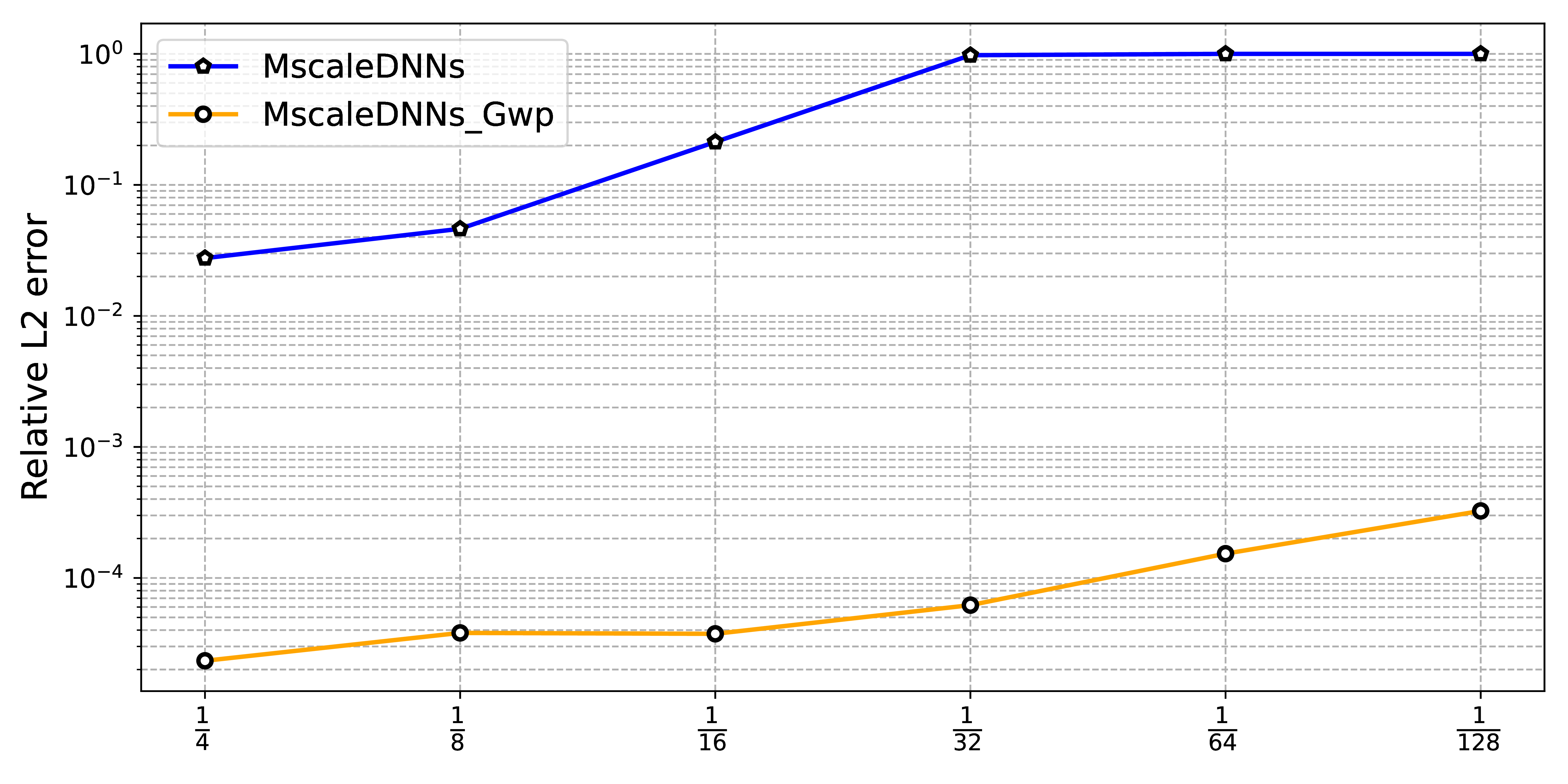}
	\caption{The comparison of the standard MscaleDNNs and MscaleDNNs with Gaussian wave packets.}
	\label{ms1}
\end{figure}

To validate the advantage of MscaleDNNs with Gaussian wave packets compared to standard MscaleDNNs, we employ standard MscaleDNNs to directly solve the Schrödinger equation (\ref{semi_sch}) under the same settings used in this subsection. As proposed by \cite{liu2020multi}, the standard MscaleDNNs architecture comprises six subnetworks, 
each taking $a_i(t, x)$ as inputs, where $a_i=2^i, $ for $i=0, 1, 2, 3, 4,$ and $ 5$. 
The non-dimensional Planck's constant is set to $\varepsilon=\frac{1}{4}, \ \frac{1}{8}, \ \frac{1}{16}, \ \frac{1}{32}, \ \frac{1}{64},  \ \frac{1}{128}$. We report the relative $L^2$ error of $\psi$ obtained using the standard MscaleDNNs in Figure \ref{ms1}. Due to the highly oscillatory nature of $\psi$ in both $x$ and $t$, the standard MscaleDNNs fail to achieve an accurate solution, even $\varepsilon=\frac{1}{32}$. This underscores the significant advantage of combining standard MscaleDNNs with Gaussian wave packets. 

\begin{table}[H]
  \begin{center}
  \caption{The harmonic potential $V(x)=\frac{1}{2} x^2$. Relative $L^2$ error of 1D examples when employing MscaleDNNs with Gaussian wave packets for the Schrödinger equation (\ref{semi_sch}) and different number of sample points (NoSP).}
  \label{NoSPtab2}
  \begin{tabular}{cccccccc}
    \hline
    NoSP & 100 & 200 & 400 & 800 & 1,600 & 3,200 & 6,400 \\
    \hline
    $\varepsilon = \frac{4}{25}$ & 1.276e-04 & 4.793e-05 & 3.716e-05 & 2.639e-05 & 4.095e-05 & 1.736e-05 & 2.962e-05 \\
    \hline
    $\varepsilon = \frac{1}{25}$ & 5.434e-04 & 1.132e-04 & 1.783e-04 & 8.685e-05 & 5.524e-05 & 2.558e-05 & 4.748e-05 \\
    \hline 
    $\varepsilon = \frac{1}{100}$ & 3.021e-03 & 4.453e-04 & 7.269e-04 & 3.065e-04 & 1.599e-04 & 1.648e-04 & 2.088e-04 \\
    \hline 
    $\varepsilon = \frac{1}{400}$ & 1.137e-02 & 1.765e-03 & 3.016e-03 & 1.198e-03 & 5.811e-04 & 5.312e-04 & 8.699e-04 \\
    \hline 
    $\varepsilon = \frac{1}{1600}$ & 4.287e-02 & 7.148e-03 & 1.239e-02 & 4.559e-03 & 2.432e-03 & 2.195e-03 & 2.468e-03 \\
    \hline 
    $\varepsilon = \frac{1}{6400}$ & 1.670e-01 & 2.908e-02 & 4.765e-02 & 1.700e-02 & 9.405e-03 & 6.795e-03 & 9.968e-03 \\
    \hline 
  \end{tabular}
  
  \end{center}
\end{table}


\begin{table}[H]
	\begin{center}
	\caption{Relative $L^2$ error of 1D examples when employ PINNs or MscaleDNNs with Gaussian wave packets for the Schrödinger equation (\ref{semi_sch}) and different potentials.}
	\begin{tabular}{ccccccc}
	  \hline
	  $\varepsilon$ & $\frac{4}{25}$ & $\frac{1}{25}$ & $\frac{1}{100}$ & $\frac{1}{400}$ & $\frac{1}{1600}$ & $\frac{1}{6400}$ \\
	  \hline 
	  PINNs for $V(x) = \frac{1}{2}x^2$ & 1.522e-03 & 4.268e-03 & 1.293e-02 & 4.904e-02 & 1.853e-01 & 6.877e-01 \\
	  \hline
	  MscaleDNNs for $V(x)=\frac{1}{2}x^2$  & 3.192e-05 & 5.512e-05 & 1.866e-04 & 5.154e-04 & 1.973e-03 & 7.956e-03\\
	  \hline
	  PINNs for $V(x) = 1 - \cos(x)$ & 4.241e-02 & 2.174e-02 & 1.065e-02 & 7.408e-03 & 1.048e-02 & 3.251e-02 \\
	  \hline
	  MscaleDNNs for $V(x) = 1 - \cos(x)$  & 4.201e-02 & 2.172e-02 & 1.092e-02 & 5.501e-03 & 3.408e-03 & 8.404e-03 \\
	  \hline
	  \label{table1}
	\end{tabular}
  \end{center}
\end{table}

\subsubsection{Torsional potential: $V(x) = 1 - \cos(x)$.}
\label{1d_tor}
Let us consider the torsional potential $V(x) = 1 - \cos(x)$ and the following initial condition:
\begin{equation*}
	\psi(x, 0) = \exp\left[\frac{\mathrm{i}}{\varepsilon}\left(\alpha(0) \left(x - q(0)\right)^2 + p(0)\left(x - q(0)\right) + \gamma(0)\right)\right],
\end{equation*}
where $q(0)=0.5\pi, \ p(0) = 0, \ \alpha(0) = i, \ \gamma_{\text{re}}(0) = 0$ 
and $\gamma_{\text{im}}(0)$ is chosen 
such that $\int_{-\infty}^{\infty} |\psi(x, 0)|\mathrm{d}x=1$.
We evaluate the accuracy of PINNs and MscaleDNNs with Gaussian wave packets for 
$\varepsilon = \frac{4}{25}, \ \frac{1}{25}, \ \frac{1}{100}, \ \frac{1}{400}, \ \frac{1}{1600}$, and $  \frac{1}{6400}$.
The relative $L^2$ errors of the two methods are reported in Table \ref{table1}. 
In the case of non-harmonic potential $V(x)$, the relative $L^2$ error 
consists of two components: the model error of Gaussian wave packets 
and the error associated with solving the ODE system (\ref{gwp_ode}). 
The model error of Gaussian wave packets decreases as $\varepsilon$ 
decreases and dominates the total relative $L^2$ error for relatively 
large values of $\varepsilon$. This is validated by the numerical 
results presented in Table \ref{table1} and explains why the two methods 
exhibit nearly identical errors for larger $\varepsilon$. However, 
for $\varepsilon<1/100$, the relative error of MscaleDNNs becomes smaller 
than that of PINNs, and this advantage further expands as $\varepsilon$ 
decreases. This demonstrates the superior capability of 
MscaleDNNs in handling highly oscillatory solutions for small 
$\varepsilon$ values in the context of non-harmonic potentials.

\subsection{2D examples}
Since $\Omega$ is a 2D domain, 
the scale-valued function $q,\, p,\,\alpha$ in Gaussian wave packets must be transitioned to vector-valued functions $\bm{q}$, $\bm{p}$, and complex-valued symmetric matrix function $A$.
For conventional DNNs with Gaussian wave packets, the input remains a single variable $t$, while the output includes multiple components corresponding to 
$\bm{q}$, $\bm{p}$, $A$, and ${\gamma}$, 
resulting in an output dimension of 12.
Consequently, the network architecture is designed as [1, 100, 400, 400, 400, 12], indicating the number of neurons from the input to the output layer.
For MscaleDNNs with Gaussian wave packets, the input consists of multiple embeddings
$[0.1t, 0.2t, \dots, 10t]$ rather than the raw time variable $t$, and 
the network is structured with layer sizes [100, 400, 400, 400, 12].
All other settings, including training and data configurations, 
remain consistent with those used in the 1D 
cases discussed in subsection \ref{1d_example}.

\subsubsection{Harmonic potential: $V(\bm{x}) = \frac{x_1^2 + x_2^2}{2}$.}
\label{2d_har}
Let us consider a harmonic potential $V(\bm{x}) = \frac{x_1^2 + x_2^2}{2}$, where $\bm{x}=(x_1, x_2)$.  
The initial condition is given by:
\begin{equation*}
	\psi(\bm{x}, 0) =  \exp\left[\frac{\mathrm{i}}{\varepsilon}\left(\left(\bm{x} - \bm{q}(0)\right)^TA(0) \left(\bm{x} - \bm{q}(0)\right) + \bm{p}(0)^T\left(\bm{x} - \bm{q}(0)\right) + {\gamma}(0)\right)\right],
\end{equation*}
where $\bm{q}(0)=(0.5, 0.5)^T, \ \bm{p}(0) = (1.0, 0.5)^T, \ A(0) = \mathrm{i}\cdot\text{diag}(1.0, 0.8)$, and $\gamma(0) = 0.75$.
The spatial domain is $\Omega = [-2,2]^2$, 
and the time horizon is set to $T=1$.
We consider different values of the non-dimensional Planck's constant,
$\varepsilon = \frac{4}{25}, \ \frac{1}{25}, \ \frac{1}{100}, \ \frac{1}{400}, \ \frac{1}{1600}$, and $ \frac{1}{6400}$.
The relative errors of the two methods are presented in Table \ref{table2}.
Since the potential is harmonic, 
the variation in relative $L^2$ error for different $\varepsilon$ follows the trends observed in the one-dimensional case discussed in subsection \ref{1d_har}.
Overall, the accuracy of MscaleDNNs surpasses that of PINNs by one to two orders of magnitude.

\begin{table}[H]
	\begin{center}
	\caption{Relative $L^2$ error of 2D examples when employ PINNs or MscaleDNNs with Gaussian wave packets for the Schrödinger equation (\ref{semi_sch})
	with $\varepsilon = \frac{4}{25}, \ \frac{1}{25}, \ \frac{1}{100}, \ \frac{1}{400}, \ \frac{1}{1600}, \ \frac{1}{6400}$ and different potentials.}
	\begin{tabular}{ccccccc}
	  \hline
	  $\varepsilon$ & $\frac{4}{25}$ & $\frac{1}{25}$ & $\frac{1}{100}$ & $\frac{1}{400}$ & $\frac{1}{1600}$ & $\frac{1}{6400}$ \\
	  \hline
	  PINNs for $V(\bm{x}) = \frac{x_1^2 + x_2^2}{2}$ & 1.598e-03 & 1.553e-03 & 3.426e-03 & 1.367e-02 & 9.400e-02 & 1.668e-01 \\
	  \hline
	  MscaleDNNs for $V(\bm{x}) = \frac{x_1^2 + x_2^2}{2}$ & 5.100e-05 & 9.683e-05 & 2.347e-04 & 6.379e-04 & 2.279e-03 & 8.793e-03 \\
	  \hline 
	  PINNs for $V(\bm{x}) = 2 - \sum_{i=1}^2\cos(x_i)$ & 4.366e-02 & 2.081e-02 & 1.145e-02 & 2.527e-02 & 9.657e-02 & 3.817e-01 \\
	  \hline 
	  MscaleDNNs for $V(\bm{x}) = 2 - \sum_{i=1}^2\cos(x_i)$ & 4.397e-02 & 2.168e-02 & 1.086e-02 & 5.807e-03 & 2.461e-03 & 9.867e-03\\
	  \hline 
	  \label{table2}
	\end{tabular}
  \end{center}
\end{table}
\subsubsection{Torsional potential: $V(\bm{x}) = 2 - \cos(x_1) - \cos(x_2)$.}
\label{2d_tor}
In this case, the potential is defined as $ 
	V(\bm{x})=2 - \cos(x_1) - \cos(x_2) $. The initial condition is specified as
$\bm{q}(0)=(1,0)^T, \ \bm{p}(0)=(0, 0)^T, \ A= \frac{\mathrm{i}}{2}\text{diag}(1,1)$. 
The real part of ${\gamma}$ is set to $0$, and the imaginary part 
of ${\gamma}$ is determined such that:
\begin{equation*}
	\int_{\mathbb{R}^2}|\psi(\bm{x},0)|^2\mathrm{d}\bm{x} = 1.
\end{equation*} 
The spatial domain is $\Omega = [-\pi,\pi]^2$, 
and the time horizon is $T=1$. We evaluate 
the performance of PINNs and
MscaleDNNs with Gaussian wave packets for 
$\varepsilon = \frac{4}{25},$ $\frac{1}{25},$ 
$\frac{1}{100},$ $\frac{1}{400},$ 
$\frac{1}{1600},$ and $\frac{1}{6400}$. 
The relative errors for the two methods are listed in Table \ref{table2}.
For this non-harmonic potential, 
the variation in relative $L^2$ error across different $\varepsilon$
values aligns with the trends observed in the 1D case discussed in subsection \ref{1d_tor}. For $\varepsilon=\frac 1 {6400}$ the relative $L^2$ error of MscaleDNNs is nearly two orders of magnitude smaller than that of PINNs.

\subsubsection{Modified Henon-Heiles potential: $V(\bm{x}) = \frac{1}{2}(x_1^2 + x_2^2)+\sigma_*(x_1x_2^2-\frac{1}{3}x_1^3) + \frac{1}{16}\sigma_*^2(x_1^2+x_2^2)^2$.}
\label{2d_hen}
A more complex potential, named as modified Henon-Heiles potential, is given by: 
\begin{equation*}
	V(\bm{x}) = \frac{1}{2}(x_1^2 + x_2^2)+\sigma_*(x_1x_2^2-\frac{1}{3}x_1^3) + \frac{1}{16}\sigma_*^2(x_1^2+x_2^2)^2,
\end{equation*}
where $\sigma_*=0.2$. 
The initial condition is a Gaussian wave packet with 
$\bm{q}(0)=(1.8,0)^T, \ \bm{p}(0)=(0, 1.2)^T,$  and $ A= \frac{\mathrm{i}}{2}\text{diag}(0.4465,1.0416)$. 
The real part of $\gamma$ is set to be $0$, and the imaginary part 
of $\gamma$ can be obtained from 
\begin{equation*}
	\int_{\mathbb{R}^2}|\psi(\bm{x},0)|^2\mathrm{d}\bm{x} = 1.
\end{equation*}
We take $\Omega = [-\pi,\pi]^2$, and $T=1$. For 
$\varepsilon =  \frac{1}{25}, $ and $ \frac{1}{100}$, MscaleDNNs with Gaussian wave packets are used to obtain a prediction solution of the Schrödinger equation. \par 
\begin{figure}[htbp]
	\centering
	\subfloat[]{\includegraphics[width=0.9\linewidth]{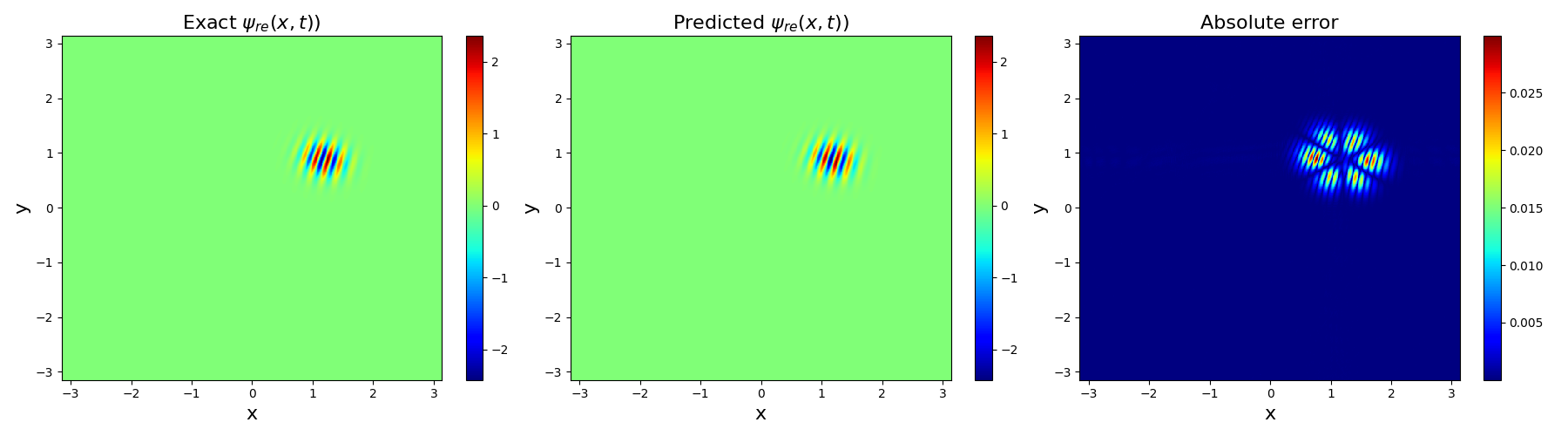}}\hfill
	\subfloat[]{\includegraphics[width=0.9\linewidth]{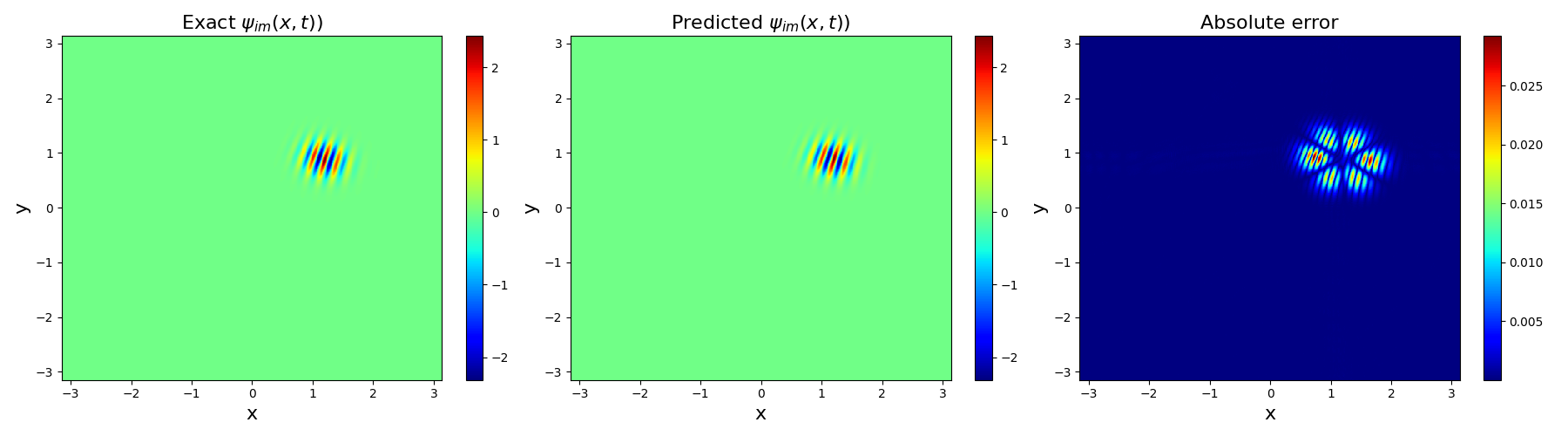}}\hfill
	\caption{Schrödinger equation near the semi-classical limit with modified Henon-Heiles potential and $\varepsilon=\frac{1}{25}$.
	(a) Left: the real part of reference solution $\psi_{\text{re}}(\bm{x},1)$. 
	Mid: the real part of prediction solution $\psi_{\text{re}}(\bm{x},1;\theta)$.
	Right: absolute error of real part $\left|\psi_{\text{re}}(\bm{x},1) - \psi_{\text{re}}(\bm{x},1;\theta)\right|$.
	(b) Left: the imaginary part of reference solution $\psi_{\text{im}}(\bm{x},1)$. 
	Mid: the imaginary part of prediction solution $\psi_{\text{im}}(\bm{x},1;\theta)$.
	Right: absolute error of the imaginary part $\left|\psi_{\text{im}}(\bm{x},1) - \psi_{\text{im}}(\bm{x},1;\theta)\right|$.}
	\label{fig:2d3}
\end{figure}
\begin{figure}[htbp]
	\centering
	\subfloat[]{\includegraphics[width=0.9\linewidth]{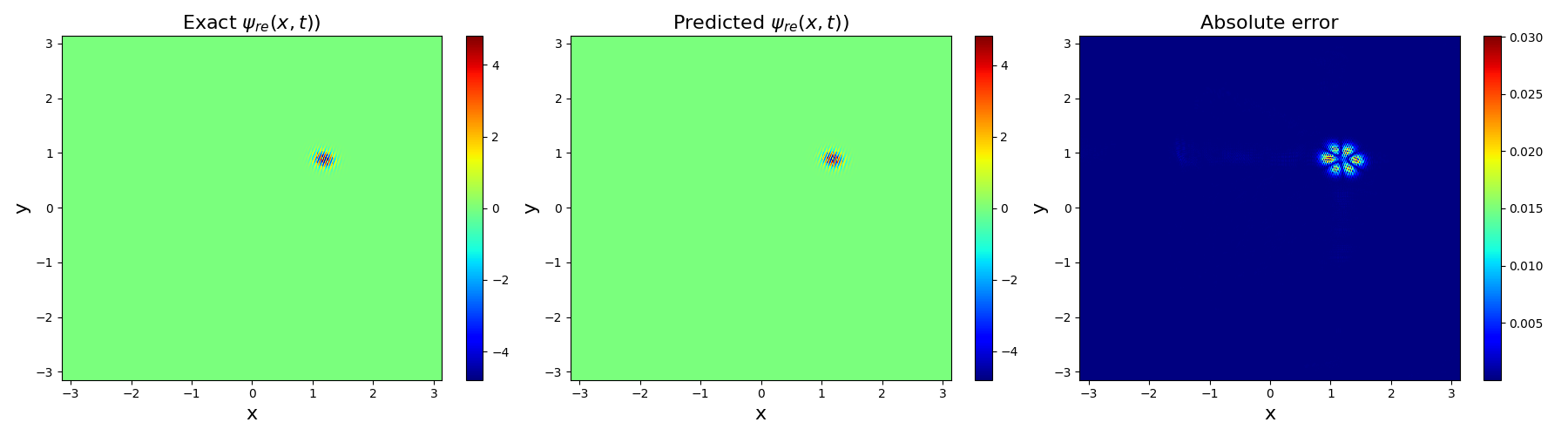}}\hfill
	\subfloat[]{\includegraphics[width=0.9\linewidth]{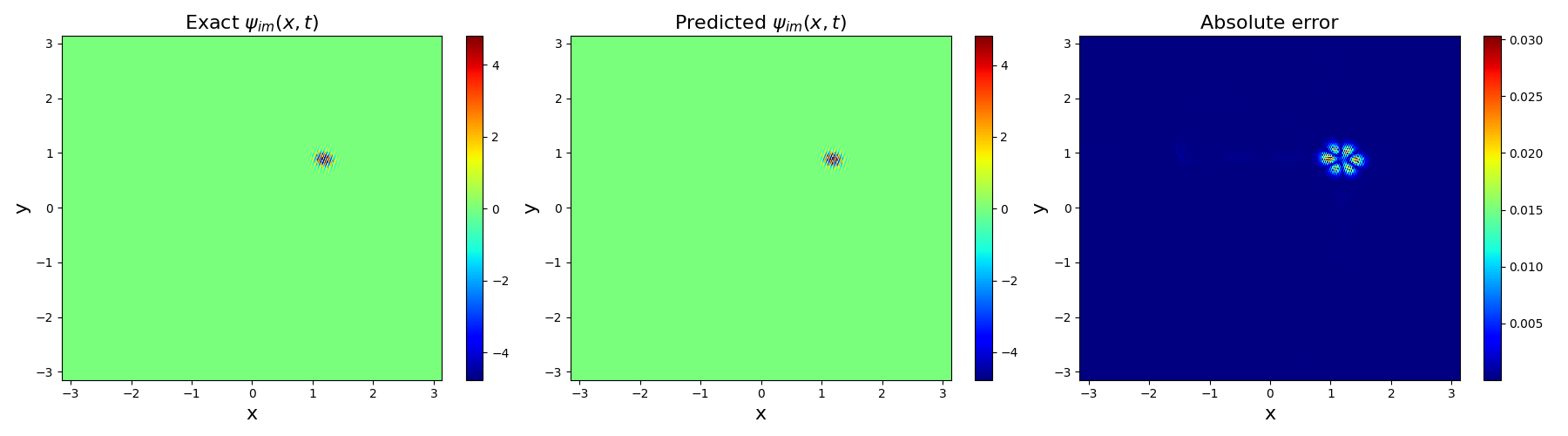}}\hfill
	\caption{Schrödinger equation near the semi-classical limit with modified Henon-Heiles potential and $\varepsilon=\frac{1}{100}$.
	(a) Left: the real part of reference solution $\psi_{\text{re}}(\bm{x},1)$. 
	Mid: the real part of prediction solution $\psi_{\text{re}}(\bm{x},1;\theta)$.
	Right: absolute error of real part $\left|\psi_{\text{re}}(\bm{x},1) - \psi_{\text{re}}(\bm{x},1;\theta)\right|$.
	(b) Left: the imaginary  part of reference solution $\psi_{\text{im}}(\bm{x},1)$. 
	Mid: the imaginary  part of prediction solution $\psi_{\text{im}}(\bm{x},1;\theta)$.
	Right: absolute error of imaginary  part $\left|\psi_{\text{im}}(\bm{x},1) - \psi_{\text{im}}(\bm{x},1;\theta)\right|$.}
	\label{fig:2d4}
\end{figure}

Figure \ref{fig:2d3} and Figure \ref{fig:2d4} show the absolute errors between 
the real and imaginary components of 
the reference solution, $\psi_{\text{re}}(\bm{x},t)$ 
and $\psi_{\text{im}}(\bm{x},t)$, and their 
approximations, $\psi_{\text{re}}(\bm{x},t;\theta)$ 
and $\psi_{\text{im}}(\bm{x},t;\theta)$, at $t=1$ 
for $\varepsilon=\frac{1}{25}$ and $\varepsilon=\frac{1}{100}$, respectively. The corresponding 
relative $L^2$ errors are 8.824e-03 and 4.515e-03. This trend demonstrates that smaller values of $\varepsilon$ lead to reduced absolute errors, indicating that the model error is the primary contributor to the overall error in this non-harmonic potential scenario. The distributions of absolute error further suggest that MscaleDNNs with Gaussian wave packets can effectively provide an accurate solution to the Schrödinger equation near the semi-classical limit. 

\subsection{4D example}
 We now consider a four dimensional example with the potential
\begin{equation*}
	V(\bm{x})=\frac{1}{2}\bm{x}^T A_p \bm{x} \ \text{~with~} \ A_p = \begin{pmatrix}
		1.0 & 0.2 & 0.2 & 0.2 \\
		0.2 & 2.0 & 0.2 & 0.2 \\
		0.2 & 0.2 & 2.0 & 0.2 \\
		0.2 & 0.2 & 0.2 & 3.0
	\end{pmatrix},
\end{equation*}
where $\bm{x}=(x_1, x_2, x_3, x_4)$.
The initial condition is set to a Gaussian wave packet with $\bm{q}(0)=(1.3, 0, -1, 0)^T, \ \bm{p}(0)=(0, 1.3, 0, 1)^T,$ and $A(0)=A_{\text{re}}(0)+iA_{\text{im}}(0)$, where 
\begin{equation*}
	A_{\text{re}}(0) = \begin{pmatrix}
		0 & 0.2 & 0 & 0\\
		0.2 & 0 & 0 & 0\\
		0 & 0 & 0 & 0\\
		0 & 0 & 0 & 0
	\end{pmatrix} \quad \text{and} \quad A_{\text{im}}(0) = \begin{pmatrix}
		1 & 0 & 0 & 0\\
		0 & 1.2 & 0 & 0\\
		0 & 0 & 0.8 & 0.3\\
		0 & 0 & 0.3 & 1.1\\
	\end{pmatrix}.
\end{equation*}
The real part of $\gamma$ is set to $0$, and the imaginary part 
of $\gamma$ is determined to satisfy the normalization condition:
\begin{equation*}
	\int_{\mathbb{R}^4}|\psi(\bm{x},0)|^2\mathrm{d}\bm{x} = 1.
\end{equation*}
Two simulations with  the non-dimensional Planck's constant, $\varepsilon=0.01$ and $\varepsilon=0.001$, are run.

In this 4D example, the matrix $A$ is 
a $4 \times 4$ complex symmetric matrix, 
necessitating output with 20 components for an effective approximation. 
To simplify the network structure, we employ four separate networks 
to independently approximate
$(\bm{q}, \bm{p})$, $A$, $\gamma_{\text{re}}$ 
and $\gamma_{\text{im}}$, respectively.
For PINNs, the architecture of the first 
network, which approximates $\bm{q}$ and $\bm{p}$, is 
configured as [1, 100, 400, 400, 400, 8], representing the number of 
neurons in each layer from input to output.
The second network, designed to approximate $A$, 
employs a fully connected architecture with layer sizes [1, 100, 100, 100, 100, 100, 20].
The remaining two networks, which 
approximate the real and imaginary parts of $\gamma$, 
share a common feedforward network 
configuration: [1, 100, 100, 100, 100, 100, 1].\par     

For MscaleDNNs, the input consists of multiple embeddings $[0.1t, 0.2t, \dots, 10t]$.
The first network, which approximates
$\bm{q}$ and $\bm{p}$, employs a network architecture 
of [100, 400, 400, 400, 8], covering the input, hidden, and output layers. 
The second network, approximating $A$, 
is structured with [100, 100, 100, 100, 100, 20] neurons across successive layers, 
while the final two networks, tasked with approximating the real and 
imaginary parts of $\gamma$, maintain the same network 
structure, with layer sizes [100, 100, 100, 100, 100, 1].
Apart from this distinction in input scaling, all other settings, 
including training parameters and data configurations, 
are consistent with those used in the 1D case discussed in subsection \ref{1d_example}.


For $\varepsilon = 0.01, $ and $ \varepsilon =0.001$, the relative $L^2$ errors for PINNs with Gaussian wave packets are 2.08e-04 and 2.07e-03, respectively. 
In contrast, MscaleDNNs Gaussian wave packets achieve relative $L^2$ errors 
of 7.48e-05 and 6.47e-04, respectively. 
These results underscore the superior performance of MscaleDNNs over PINNs in solving the 
ODE system, leading to an improvement in the accuracy of the final solution by two orders of magnitude.

\section{Numerical experiments for Physics-Informed DeepONets with Gaussian wave packets}
\label{sec5}
To demonstrate the effectiveness of physics-informed DeepONets 
with Gaussian wave packets, we conduct 
a series of numerical
experiments to solve the Schrödinger 
equation (\ref{semi_sch}) 
under various Gaussian initial conditions.
The initial conditions are modeled as:
\begin{equation}
	\label{op_ini}
	s(\bm{x}) = \exp\left(\frac{\mathrm{i}}{\epsilon}\left(A_0(\bm{x}-\bm{q}_0)^2+\bm{p_0}(\bm{x}-\bm{q}_0)+\gamma_0\right)\right)
\end{equation}
where $\bm{q}_0$, $\bm{p}_0$, $A_0$, and $\gamma_0$ 
are random variables sampled from predefined distributions.
Throughout this section, we consistently apply
the same activation function and 
network initialization as 
described in section \ref{num_exp}.
The networks are trained using mini-batch stochastic 
gradient descent, optimized with 
the Adam algorithm \cite{kingma2014adam} under default settings. 
Specifically, we use a batch size of 1,000 and 
incorporate exponential learning rate decay 
with a decay rate of 0.9 every 4,000 training iterations.\par 

We evaluate the performance of the physics-informed DeepONets using the 
mean and standard deviation of 
the relative $L^2$ error, defined as:
\begin{equation*}
	\text{mean}(E_{\text{rel}}) = \frac{1}{N} \sum_{n=1}^N E_{\text{rel}}(\psi^n_{\text{pred}},\psi^n_{\text{ref}}),
\end{equation*}
and 
\begin{equation*}
	\text{std}(E_{\text{rel}}) =\sqrt{\frac{1}{N} \sum_{n=1}^N  \left(E_{\text{rel}}(\psi^n_{\text{pred}},\psi^n_{\text{ref}}) - \text{mean}(E_{\text{rel}})\right)^2},
\end{equation*}
where 
\begin{equation*}
    E_{\text{rel}}(\psi^n_{\text{pred}},\psi^n_{\text{ref}}) = \frac{\|\psi^n_{\text{pred}} - \psi^n_{ref} \|_{L^2(\Omega\times [0,T])}}{\| \psi^n_{ref} \|_{L^2(\Omega\times [0,T])}} = \frac{\sqrt{\int_{\Omega\times [0, T]} |\psi^n_{\text{pred}}(\bm{x},t) - \psi^n_{\text{ref}}(\bm{x},t)|^2 \, \mathrm{d}\bm{x}\mathrm{d}t}}{\sqrt{\int_{\Omega\times[0,T]} |\psi^n_{\text{ref}}(\bm{x},t)|^2 \, \mathrm{d}\bm{x}\mathrm{d}t}}.
\end{equation*}
The neural network outputs include
$\bm{q}(t)$, $\bm{p}(t)$, $A(t)$ and $\gamma(t)$. 
To assess their precision, we calculate the mean and
variance of their relative $L^2$ errors
using definitions analogous to those applied for $\psi$. 
These metrics offer a thorough assessment of the accuracy achieved in solving the Schrödinger equation. 
In this section, we primarily focus on the performance of physics-informed DeepONets with Gaussian wave packets. An extension to more general initial conditions using Hagedorn wave packets is provided in \ref{ha_wave_packets}.
Furthermore, an extension of DeepONets to handle 
WKB-type initial conditions using the Gaussian beam method is 
provided in \ref{gbd_wkb}.

\subsection{1D example with torsional potential: $V(x) = 1 - \cos(x)$.}
\label{op_tor_1d}
Both \cite{russo2013gaussian} and \cite{faou2009computing} 
present numerical examples for the potential $V(x) = 1 - \cos(x)$, 
where the initial values of $q(0)$ in (\ref{op_ini}) 
are $\frac{\pi}{2}$ and $1$, respectively. 
Consequently, in this example, 
we consider $q(0) \sim \mathcal{U}[0.8, 1.8]$, where
$\mathcal{U}$ denotes a uniform distribution.
This choice allows us to simultaneously 
obtain final solutions for
$q(0) = \frac{\pi}{2}$ and $1$, 
as well as for other initial values of $q(0)$ 
within the interval $[0.8, 1.8]$.
In general, we assume $T=1$ with the following initial conditions:
\begin{equation*}
	p(0), \alpha_{\text{re}}(0) \sim \mathcal{U}[-0.5,0.5], \,\,\quad \alpha_{\text{im}}(0) \sim \mathcal{U}[0.5,1.5] ,\ \,\,\quad \gamma_{\text{re}}(0)=0,
\end{equation*}
and $\gamma_{\text{im}}(0)$ chosen such that 
$\int_{-\infty}^{\infty} |\psi(x, 0)|\mathrm{d}x=1$. The non-dimensional Planck's constant $\varepsilon=0.01$.
The goal is to learn the operator that maps 
$\bm{y}(0)$ in (\ref{1dvector_gwp}) to the 
solution $\bm{y}(t)$ where $t\in [0, 1]$,
without relying on any paired input-output data.
To achieve this, we represent the operator using DeepONets
$\mathcal{G}'_{\theta}$ as defined in 
formula (\ref{multi_deeponet}), where 
$I=6$ and $J_1 = \dots = J_I = 100$.
The DeepONet comprises a branch network with 
architecture [6, 100, 100, 100, 100, 600], and a trunk network 
configured as [1, 100, 100, 100, 100, 100], where the numbers indicate the 
neurons in each layer.

\begin{figure}[htbp]
	\centering
	\subfloat[]{\includegraphics[width=0.9\linewidth]{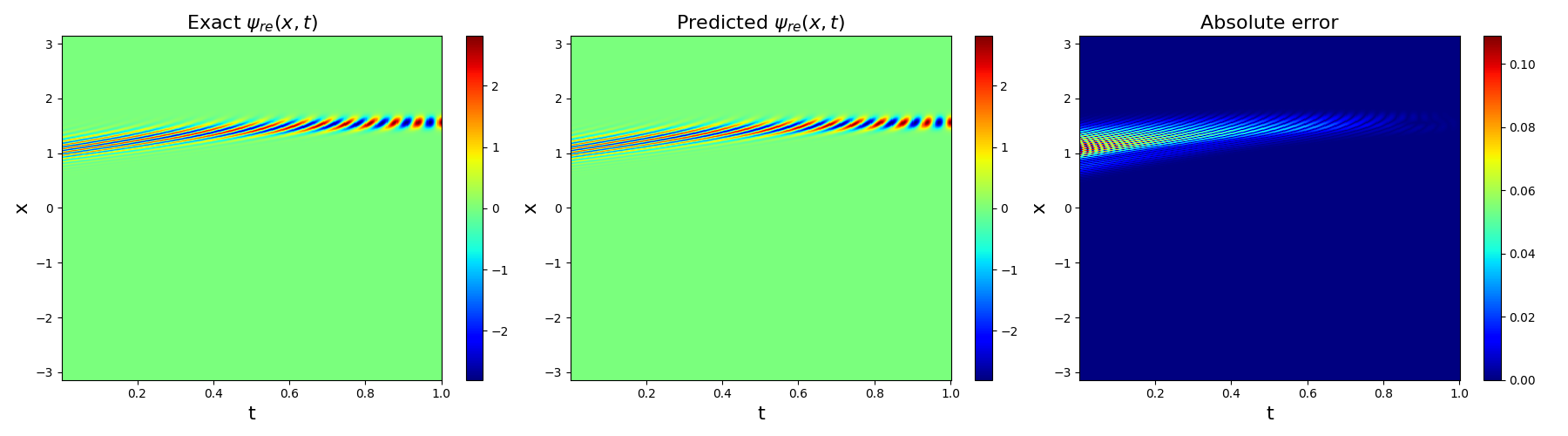}}\hfill
	\subfloat[]{\includegraphics[width=0.9\linewidth]{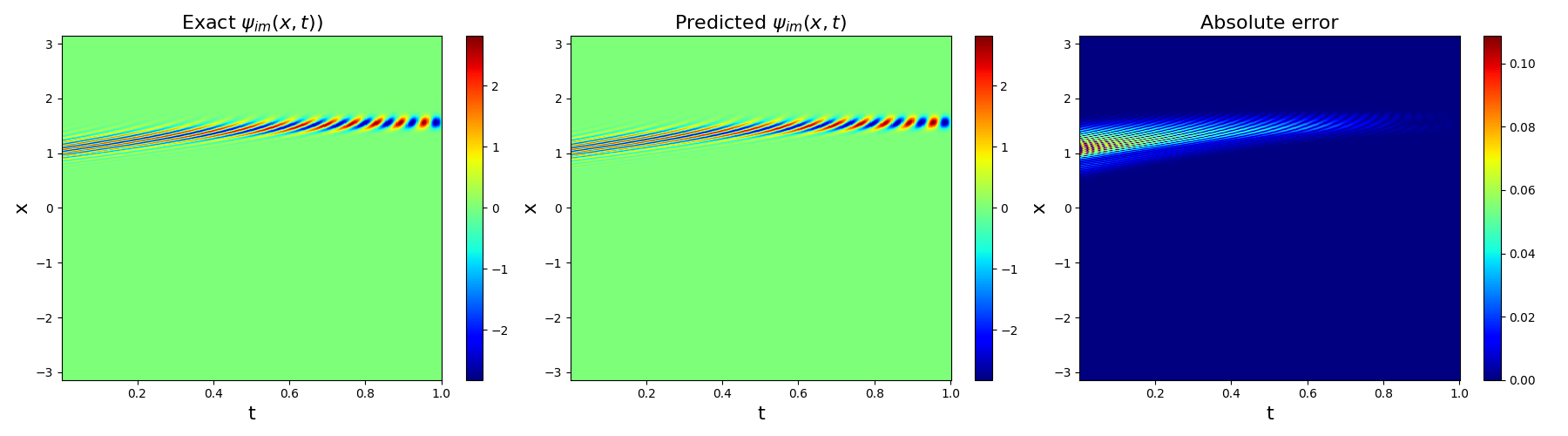}}\hfill
	\caption{1D Schrödinger equation (\ref{semi_sch}) with $\varepsilon=0.01$,  torsional potentia $V(x)=1-\cos(x)$, and 
	$q(0)=\frac{\pi}{2}, \ p(0)=0, \ \alpha(0)=\mathrm{i}, \ \gamma(0)= -\frac{1}{4}\log(\frac{2}{\pi\varepsilon})\mathrm{i}$.
	(a) Left: the real part of reference solution $\psi_{\text{re}}(x,t)$. 
	Mid: the real part of prediction solution $\psi_{\text{re}}(x,t;\theta)$.
	Right: absolute error of real part $\left|\psi_{\text{re}}(x,t) - \psi_{\text{re}}(x,t;\theta)\right|$.
	(b) Left: the imaginary  part of reference solution $\psi_{\text{im}}(x,t)$. 
	Mid: the imaginary  part of prediction solution $\psi_{\text{im}}(x,t;\theta)$.
	Right: absolute error of imaginary  part $\left|\psi_{\text{im}}(x,t) - \psi_{\text{im}}(x,t;\theta)\right|$.}
	\label{op_tor_1}
\end{figure}

\begin{figure}[htbp]
	\centering
	\subfloat[]{\includegraphics[width=0.9\linewidth]{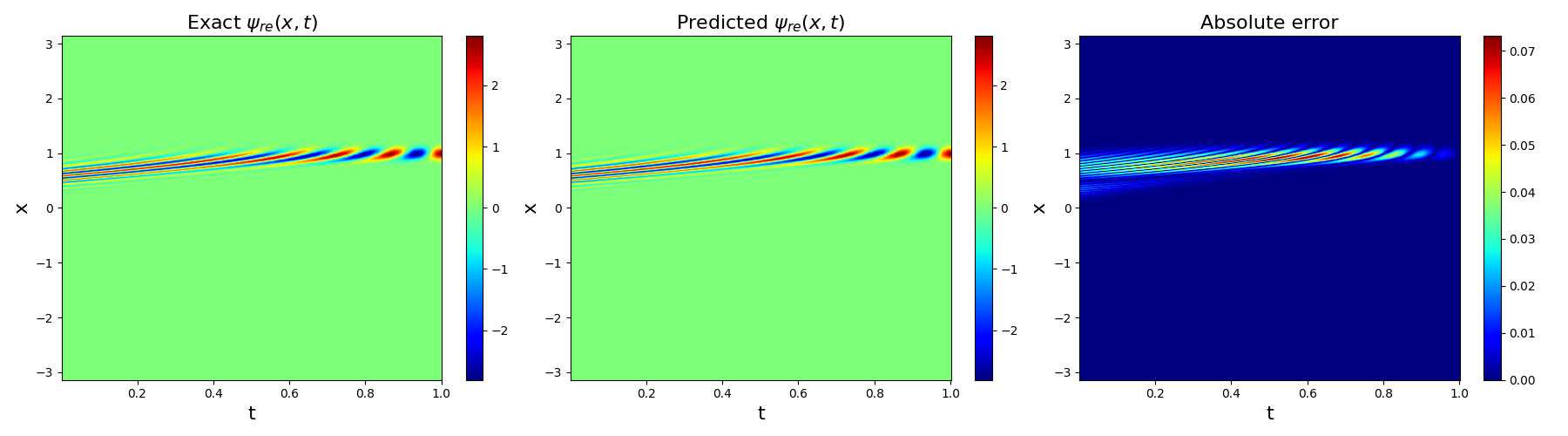}}\hfill
	\subfloat[]{\includegraphics[width=0.9\linewidth]{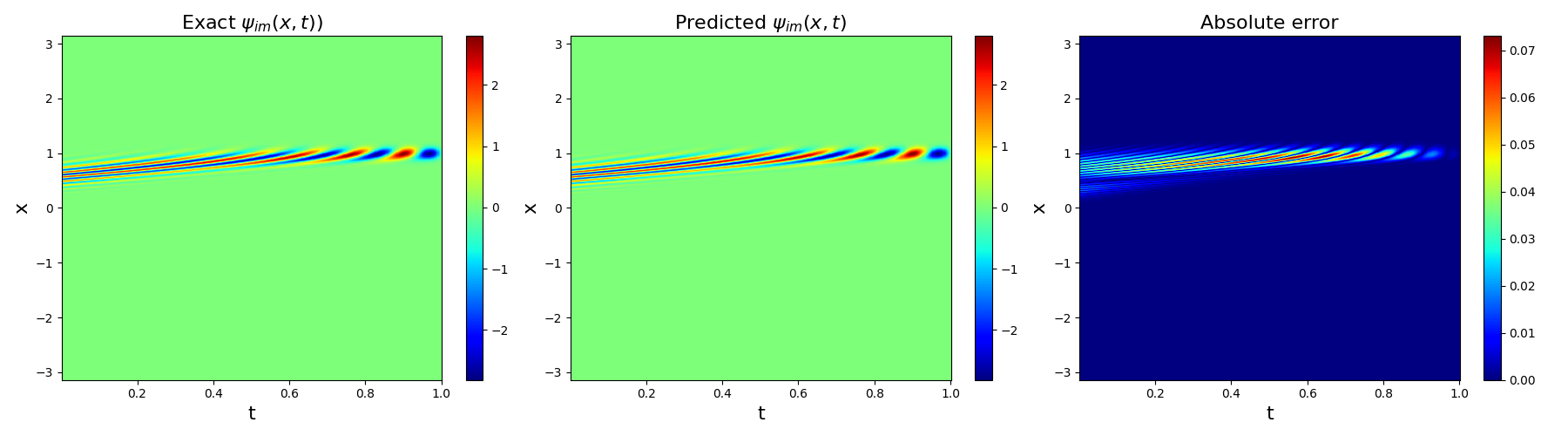}}\hfill
	\caption{1D Schrödinger equation (\ref{semi_sch}) with  $\varepsilon=0.01$, torsional potential $V(x)=1-\cos(x)$, and 
	$q(0)=1, \ p(0)=0, \ \alpha(0)=\mathrm{i}, \ \gamma(0)= -\frac{1}{4}\log(\frac{2}{\pi\varepsilon})\mathrm{i}$.
	(a) Left: the real part of reference solution $\psi_{\text{re}}(x,t)$. 
	Mid: the real part of prediction solution $\psi_{\text{re}}(x,t;\theta)$.
	Right: absolute error of real part $\left|\psi_{\text{re}}(x,t) - \psi_{\text{re}}(x,t;\theta)\right|$.
	(b) Left: the imaginary part of reference solution $\psi_{\text{im}}(x,t)$. 
	Mid: the imaginary part of prediction solution $\psi_{\text{im}}(x,t;\theta)$.
	Right: absolute error of imaginary part $\left|\psi_{\text{im}}(x,t) - \psi_{\text{im}}(x,t;\theta)\right|$.}
	\label{op_tor_2}
\end{figure}

\begin{figure}[htbp]
	\centering
	\subfloat[]{\includegraphics[width=0.9\linewidth]{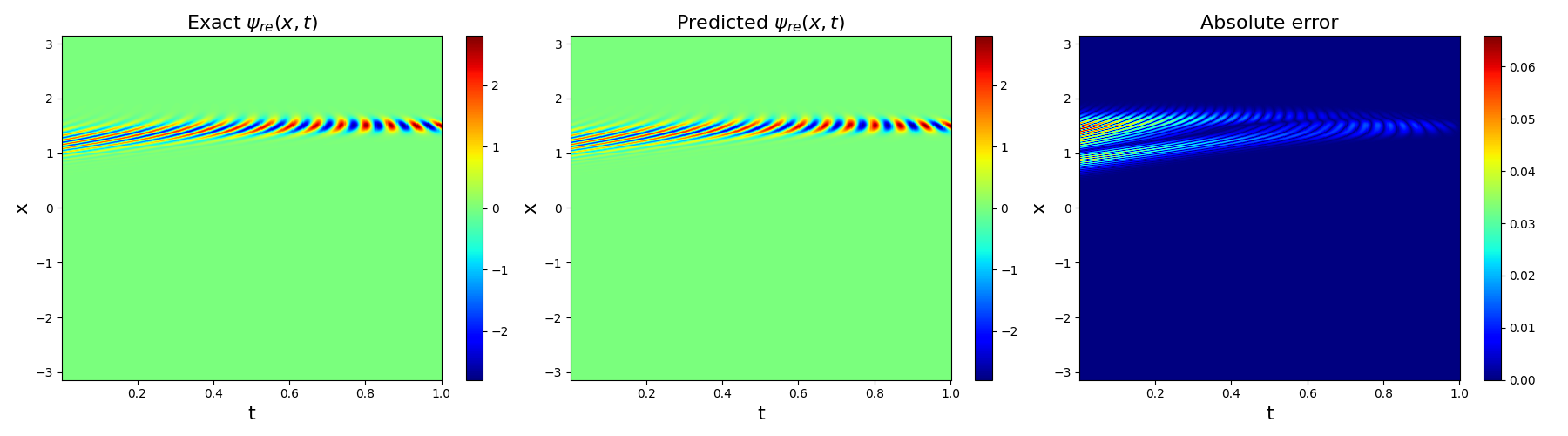}}\hfill
	\subfloat[]{\includegraphics[width=0.9\linewidth]{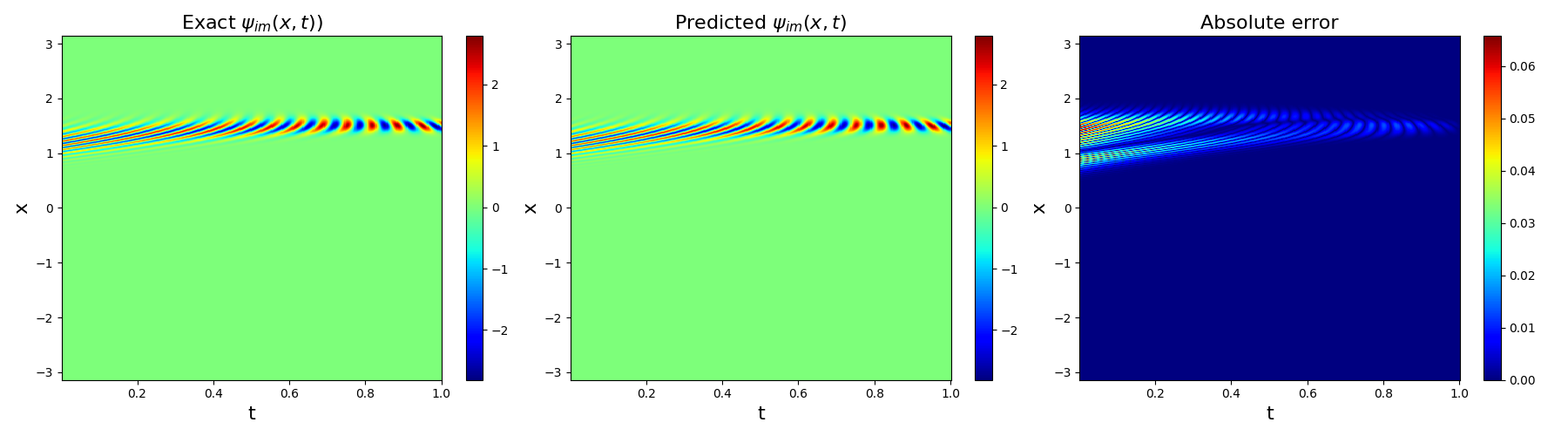}}\hfill
	\caption{1D Schrödinger equation (\ref{semi_sch}) with  $\varepsilon=0.01$, torsional potential $V(x)=1-\cos(x)$, and 
	$q(0)=1.5, \ p(0)=0.2, \ \alpha(0)=0.2 + \mathrm{i}, \ \gamma(0)= -\frac{1}{4}\log(\frac{2}{\pi\varepsilon})\mathrm{i}$.
	(a) Left: the real part of reference solution $\psi_{\text{re}}(x,t)$. 
	Mid: the real part of prediction solution $\psi_{\text{re}}(x,t;\theta)$.
	Right: absolute error of real part $\left|\psi_{\text{re}}(x,t) - \psi_{\text{re}}(x,t;\theta)\right|$.
	(b) Left: the imaginary part of reference solution $\psi_{\text{im}}(x,t)$. 
	Mid: the imaginary part of prediction solution $\psi_{\text{im}}(x,t;\theta)$.
	Right: absolute error of imaginary part $\left|\psi_{\text{im}}(x,t) - \psi_{\text{im}}(x,t;\theta)\right|$. }
	\label{app_op_tor_1}
\end{figure}

We sample $N=2,000$ initial functions 
as defined in equation (\ref{op_ini}) with random values of $q(0)$, $p(0)$, and $\alpha(0)$ drawn from the specified distributions.
In the loss function (\ref{phy_deeponet_loss}), we set $Q=500$ and randomly sample 
$\{t_j\}_{j=1}^Q$ from the interval $[0,1]$. 
We then train the physics-informed DeepONets with Gaussian 
wave packets by minimizing the
loss function (\ref{phy_deeponet_loss}) over 200,000 iterations. 
The numerical results for $q(0)=\frac{\pi}{2}$, $q(0)=1$, and  $q(0)=1.5$, are presented in Figure \ref{op_tor_1}, Figure \ref{op_tor_2}, and Figure \ref{app_op_tor_1}, respectively.
The relative $L^2$ errors for these cases are 2.69e-02, 2.39e-02, and 1.67e-02.
As evident from the results, the predictions generated by the physics-informed DeepONets exhibit strong agreement with the reference solutions.
Finally, we sample $N_{\text{test}}=100$ 
initial functions
from the distribution to compute the 
mean and standard deviation of the relative 
$L^2$ error, which are reported as $\text{mean}(E_{\text{rel}})$=2.44e-02  and $ \text{std}(E_{\text{rel}})=$  1.54e-02.

We compare the computational time of the trained neural network 
operator with that of the classical fourth-order 
Runge--Kutta (RK4) method for solving the same number 
of initial value problems using the same computational resources. Only the online inference time is considered for DeepONets, in line with the evaluation methodology adopted in \cite{li2020fourier, molinaro2023neural}. To ensure a fair comparison, the RK4 time step is set to $\Delta t = 0.1$ such that its relative error $L^2$ is consistent with that of DeepONets. Moreover, DeepONets generate predictions at the same mesh points used in the RK4 scheme. Table~\ref{ex1_compare_t} reports the inference time of DeepONets and the simulation time of the RK4 method for
$N_{\text{test}}=10,000, \ 20,000, \ 40,000$ test cases. As the number of initial conditions increases, the prediction time of DeepONets grows only marginally, demonstrating significantly more favorable scaling compared to the RK4 method. Notably, for $N_{\text{test}} \geq 10{,}000$, DeepONets are at least twice as fast as RK4.
An additional advantage of DeepONets is their ability to generate solutions at arbitrary time points without the need for interpolation. Due to this property, the inference time of DeepONets increases only marginally as the number of mesh points grows. For instance, generating solutions at 100 mesh points for $N_{\text{test}} = 10{,}000$ test cases takes only 4.58 seconds. In contrast, the RK4 scheme would either require costly recomputation with a smaller time step
$\Delta t$ or interpolation of coarse-grained solutions. For example, the simulation time of the RK4 scheme with $\Delta t=0.01$ and $N_{\text{test}}=10,000$ is 67.85 seconds. This further highlights the computational advantages of the proposed neural operator framework.

\begin{table}[htbp]
  \begin{center}

  \caption{Runtime comparison of RK4 and DeepONets for different 
  $N_{\text{test}}$ initial conditions when $V(x)=1 - \cos(x)$.}
  \begin{tabular}{cccc}
    \hline
    $N_{\text{test}}$ & $10,000$ & $20,000$ & $40,000$ \\
    \hline 
     RK4 & 7.29s & 14.06s & 28.60s \\
    \hline
     DeepONets & 3.17s & 3.28s & 3.70s \\
    \hline 
  \end{tabular}
  \label{ex1_compare_t}
  
  \end{center}
\end{table}

\subsection{1D example with harmonic potential: $V(x) = \frac{1}{2} x^2$.}
\label{op_1d_har}
\begin{figure}[htbp]
	\centering
	\subfloat[]{\includegraphics[width=0.9\linewidth]{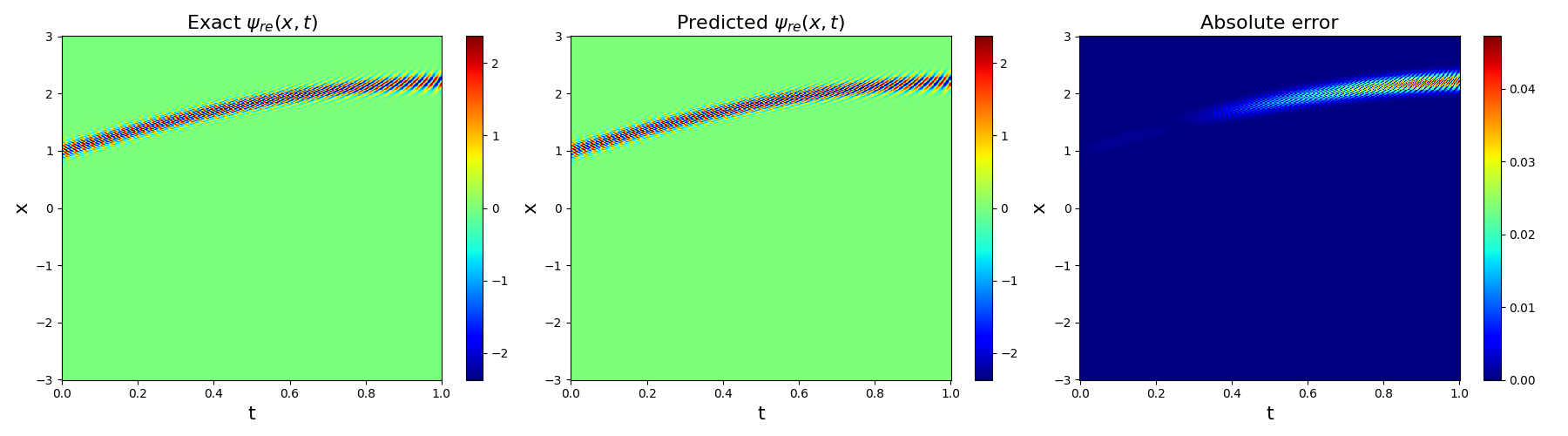}}\hfill
	\subfloat[]{\includegraphics[width=0.9\linewidth]{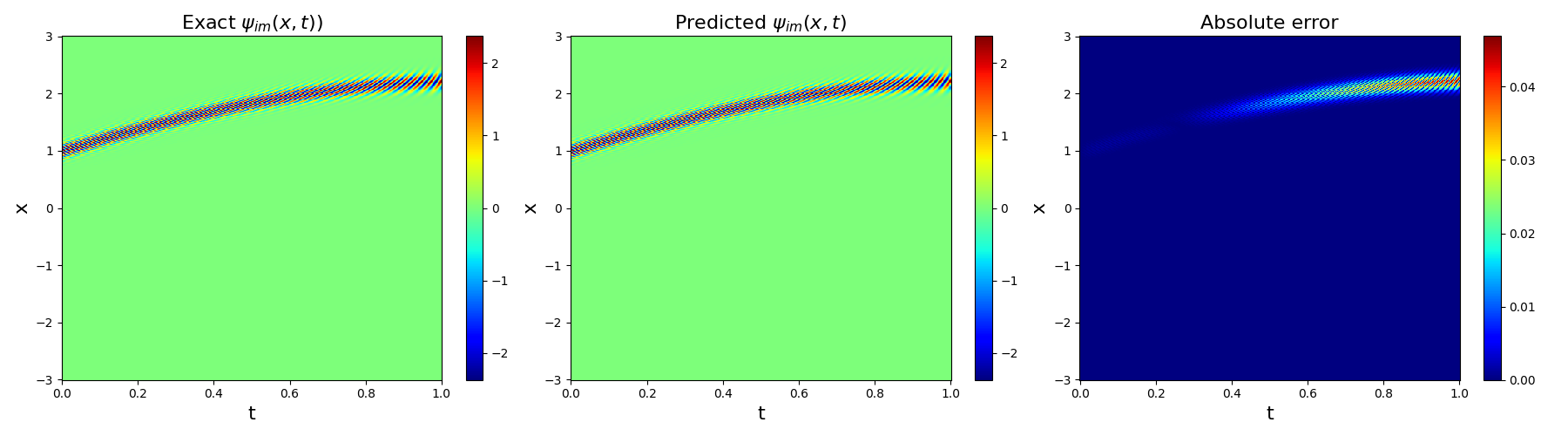}}\hfill
	\caption{1D Schrödinger equation (\ref{semi_sch}) with $\varepsilon=0.01$, harmonic potential $V(x)=\frac{x^2}{2}$, and 
	$q(0)=1, \ p(0)=2, \ \alpha(0)=\frac{\mathrm{i}}{2}, \ \gamma(0)= -\frac{1}{4}\log(\frac{1}{\pi\varepsilon})\mathrm{i}$:
	(a) Left: the real part of reference solution $\psi_{\text{re}}(x,t)$. 
	Mid: the real part of prediction solution $\psi_{\text{re}}(x,t;\theta)$.
	Right: absolute error of real part $\left|\psi_{\text{re}}(x,t) - \psi_{\text{re}}(x,t;\theta)\right|$.
	(b) Left: the imaginary part of reference solution $\psi_{\text{im}}(x,t)$. 
	Mid: the imaginary part of prediction solution $\psi_{\text{im}}(x,t;\theta)$.
	Right: absolute error of imaginary part $\left|\psi_{\text{im}}(x,t) - \psi_{\text{im}}(x,t;\theta)\right|$.}
	\label{op_har_1d1}
\end{figure}

\begin{figure}[htbp]
	\centering
	\subfloat[]{\includegraphics[width=0.9\linewidth]{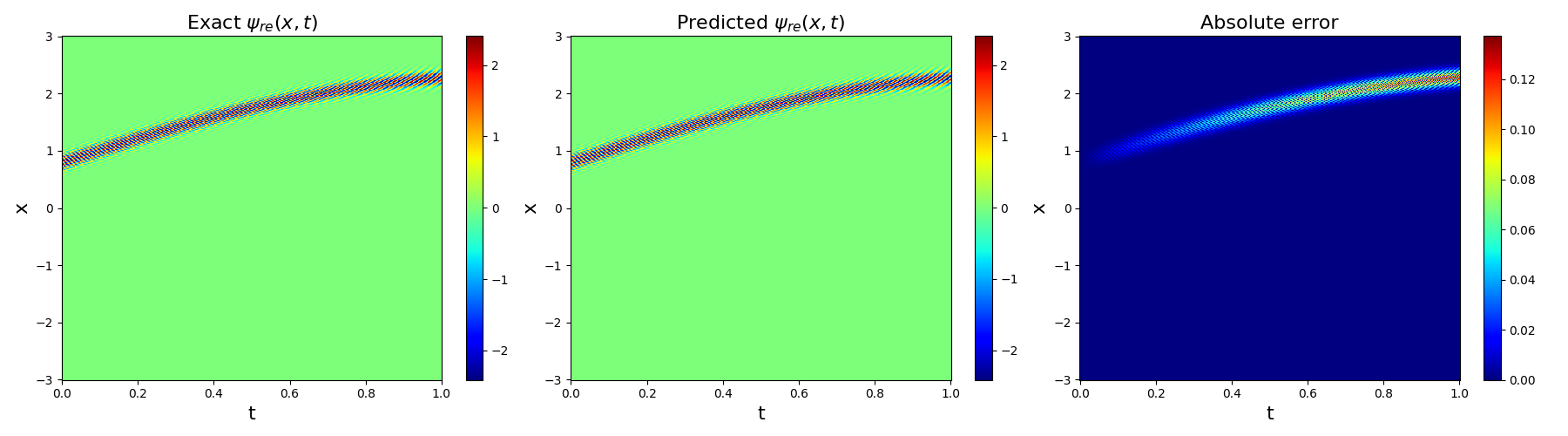}}\hfill
	\subfloat[]{\includegraphics[width=0.9\linewidth]{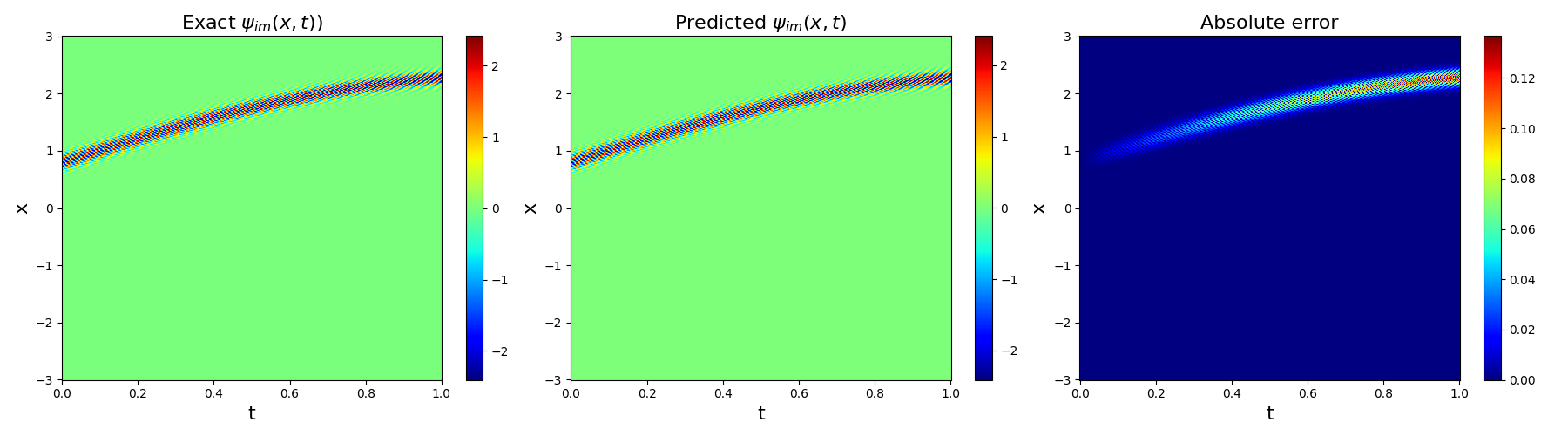}}\hfill
	\caption{1D Schrödinger equation (\ref{semi_sch}) with $\varepsilon=0.01$, harmonic potential $V(x)=\frac{x^2}{2}$, and 
	$q(0)=0.8, \ p(0)=2.2, \ \alpha(0)=\frac{2\mathrm{i}}{5}, \ \gamma(0)= -\frac{1}{4}\log(\frac{4}{5\pi\varepsilon})\mathrm{i}$:
	(a) Left: the real part of reference solution $\psi_{\text{re}}(x,t)$. 
	Mid: the real part of prediction solution $\psi_{\text{re}}(x,t;\theta)$.
	Right: absolute error of real part $\left|\psi_{\text{re}}(x,t) - \psi_{\text{re}}(x,t;\theta)\right|$.
	(b) Left: the imaginary part of reference solution $\psi_{\text{im}}(x,t)$. 
	Mid: the imaginary part of prediction solution $\psi_{\text{im}}(x,t;\theta)$.
	Right: absolute error of imaginary part $\left|\psi_{\text{im}}(x,t) - \psi_{\text{im}}(x,t;\theta)\right|$. The relative $L^2$ error of $\psi$ is 3.60e-02.}
	\label{op_har_1d2}
\end{figure}

This 1D example involves a 
harmonic potential $V(x) = \frac{1}{2} x^2$. We assume the initial conditions $q(0)$, $p(0)$, and $\alpha_{\text{im}}(0)$ satisfy the following distributions:
\begin{equation*}
	q(0) \sim \mathcal{U}[0.5, 1.5], \,\quad p(0) \sim \mathcal{U}[1.5, 2.5], \,\quad \alpha_{\text{im}}(0) \sim \mathcal{U}[0.2, 0.7].  
\end{equation*}
The initial conditions $\alpha_{\text{re}}(0)=\gamma_{\text{re}}(0)=0$, and $\gamma_{\text{im}}(0)$
is chosen such that $\int_{-\infty}^{\infty} |\psi(x, 0)|\mathrm{d}x=1$. The non-dimensional Planck's constant is set to $\varepsilon=0.01$.
The structure of the DeepONets, the training points, and the training strategies remain unchanged from those used in subsection \ref{op_tor_1d}.    
We sample $N_{\text{test}}=1,000$ 
initial functions as defined in equation (\ref{op_ini}) with random values of $q(0)$, $p(0)$, and $\alpha_{\text{im}}$ 
drawn from the specified distributions. The distributions of $\psi$ obtained by physics-informed DeepONets are 
displayed in Figure \ref{op_har_1d1} and Figure \ref{op_har_1d2}, which demonstrate excellent agreement 
with the reference solutions. 
The mean and standard deviation of the relative 
$L^2$ error are 1.93e-02 and 1.60e-02, respectively. 

In the case of harmonic potentials, the Gaussian wave packet exhibits no model error. The mean and standard deviation of relative $L^2$ error of
$q, p, \alpha, \gamma$, and the
final solution $\psi$ 
are summarized in Table \ref{table3}.
Notably, for $\varepsilon=0.01$, the mean and standard 
deviation of relative $L^2$ error of the final solution 
are precisely two orders of magnitude higher than those of the ODE system solution, as previously analyzed. 
We retain a similar experimental setup as in section~\ref{op_tor_1d} to compare the 
computational time required to solve varying numbers of initial 
value problems, denoted by $N_{\text{test}}$. 
The results, summarized in Table~\ref{ex2_compare_t}, demonstrate 
that DeepONets achieve solution speeds at least two to three 
times faster than the classical RK4 method. Moreover, while 
the computational time of RK4 increases linearly with $N_{\text{test}}$, 
the runtime of DeepONet remains nearly constant. These results further 
highlight the efficiency and scalability of the proposed neural 
operator approach.

\begin{table}[htbp]
  \begin{center}
  \caption{Runtime comparison of RK4 and DeepONets for different 
  $N_{\text{test}}$ initial conditions when $V(x)=\frac{1}{2}x^2$.}
  \begin{tabular}{cccc}
    \hline
    $N_{\text{test}}$ & $10,000$ & $20,000$ & $40,000$ \\
    \hline 
     RK4 & 11.36s & 22.61s & 45.63s \\
    \hline
     DeepONets & 4.15s & 4.72s & 5.44s \\
    \hline 
  \end{tabular}
  \label{ex2_compare_t}
  \end{center}
\end{table}

\begin{table}[htbp]
	\begin{center}
	\caption{The mean and strandard variance of relative $L^2$ error for $q, p, \alpha, \gamma$, and the
final solution $\psi$.}
	\begin{tabular}{cccccc}
	  \hline
	   & $q$ & $p$ & $\alpha$ & $\gamma$ & $\psi$ \\
	  \hline 
	  mean& 5.33e-05 & 6.95e-05 & 4.04e-04  & 1.85e-04 & 1.93e-02\\
	  \hline
	  strandard variance &4.86e-05 & 7.08e-05 & 3.40e-04 & 1.29e-04 & 1.60e-02\\
	  \hline 
	  \label{table3}
	\end{tabular}
  \end{center}
\end{table}

\subsection{2D example with harmonic potential: $V(\bm{x}) = \frac{x_1^2 + x_2^2}{2}$.}
We consider a 2D example with the harmonic potential 
$V(\bm{x}) = \frac{x_1^2 + x_2^2}{2}$. The
initial functions are given by equation (\ref{op_ini}),
with $\bm{q}(0)$ and $\bm{p}(0)$ satisfying
\begin{equation*}
	\bm{q}(0) \sim \mathcal{U}[0.2,1.2]^2, \ \quad \bm{p}(0) \sim \mathcal{U}[0.2,1.2]^2.
\end{equation*}
The matrix $A(0)$ is a complex symmetric matrix of the following form
\begin{equation*}
	A(0) = \begin{bmatrix}
		\alpha_{11}(0) & \alpha_{12}(0) \\
		\alpha_{12}(0) & \alpha_{22}(0) \\
		\end{bmatrix}.
\end{equation*}
We assume $\gamma(0)=0.75$, $A_{\text{re}}(0) = 0$, $\alpha_{12, \text{im}}(0)=0$, and
\begin{equation*}
	(\alpha_{11, \text{im}}(0),\alpha_{22, \text{im}}(0)) \sim \mathcal{U}[0.2,1.2]^2.
\end{equation*} 
In this simulation, we apply the operator 
$\mathcal{G}'_{\theta}$, as defined in formula (\ref{multi_deeponet}) 
with $I=12,$ and $ J_1 = \dots = J_I = 50$. 
The architecture of the branch network is defined 
as [12, 600, 600, 600, 600, 600], indicating one input layer with 
12 neurons followed by five hidden layers. The trunk network follows 
a structure of [1, 100, 100, 100, 100, 50], with a single input and 
progressively deepening hidden layers.

\begin{figure}[htbp]
	\centering
	\subfloat[]{\includegraphics[width=0.9\linewidth]{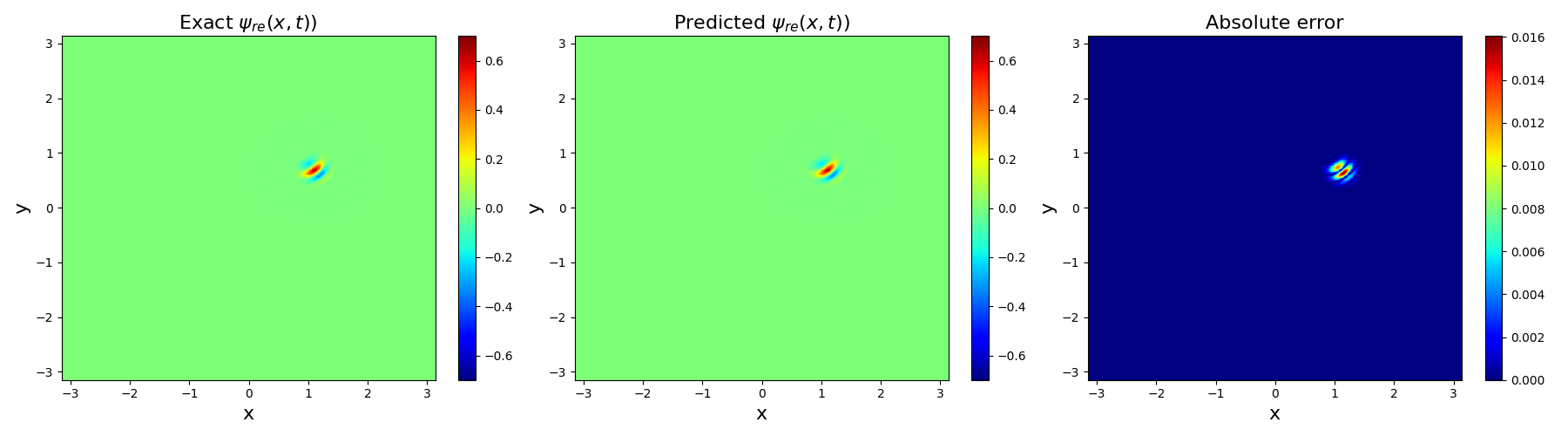}}\hfill
	\subfloat[]{\includegraphics[width=0.9\linewidth]{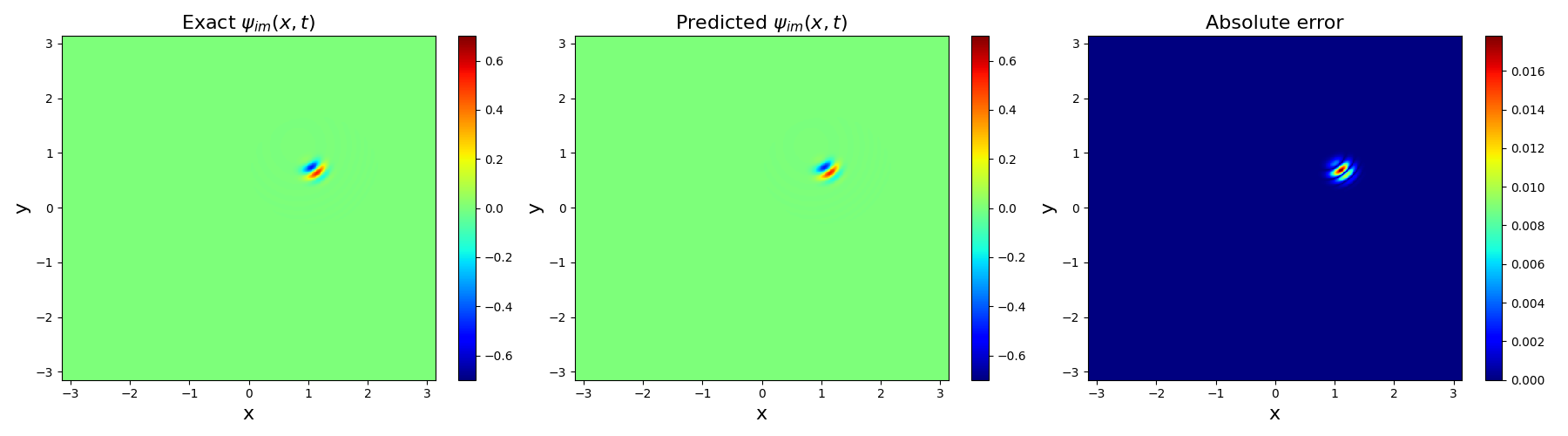}}\hfill
	\caption{2D Schrödinger equation with $\varepsilon=0.01$, harmonic potential $V(\bm{x}) = \frac{x_1^2 + x_2^2}{2}$, and 
	$\bm{q}(0)=(0.5, 0.5)^T, \ \bm{p}(0) = (1.0, 0.5)^T, \ A(0) = \mathrm{i}\text{diag}(1.0, 0.8), \ \gamma(0) = 0.75$.
	(a) Left: the real part of reference solution $\psi_{\text{re}}(\bm{x},1)$. 
	Mid: the real part of prediction solution $\psi_{\text{re}}(\bm{x},1;\theta)$.
	Right: absolute error of real part $\left|\psi_{\text{re}}(\bm{x},1) - \psi_{\text{re}}(\bm{x},1;\theta)\right|$.
	(b) Left: the imaginary part of reference solution $\psi_{\text{im}}(\bm{x},1)$. 
	Mid: the imaginary part of prediction solution $\psi_{\text{im}}(\bm{x},1;\theta)$.
	Right: absolute error of imaginary part $\left|\psi_{\text{im}}(\bm{x},1) - \psi_{\text{im}}(\bm{x},1;\theta)\right|$.}
	\label{op_2d_1}
\end{figure}


\begin{figure}[htbp]
	\centering
	\subfloat[]{\includegraphics[width=0.9\linewidth]{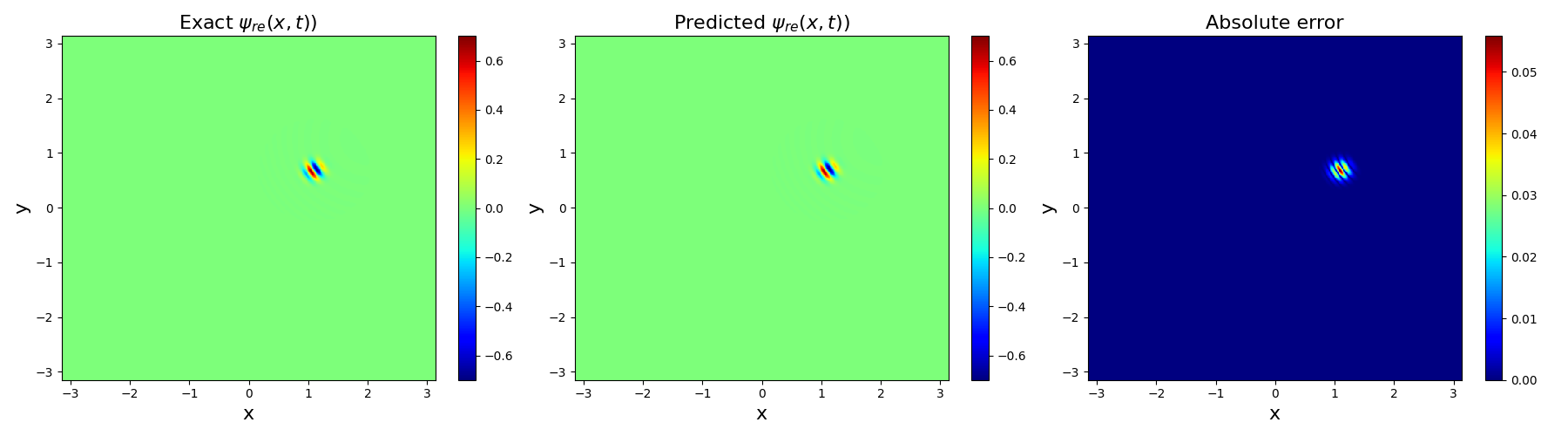}}\hfill
	\subfloat[]{\includegraphics[width=0.9\linewidth]{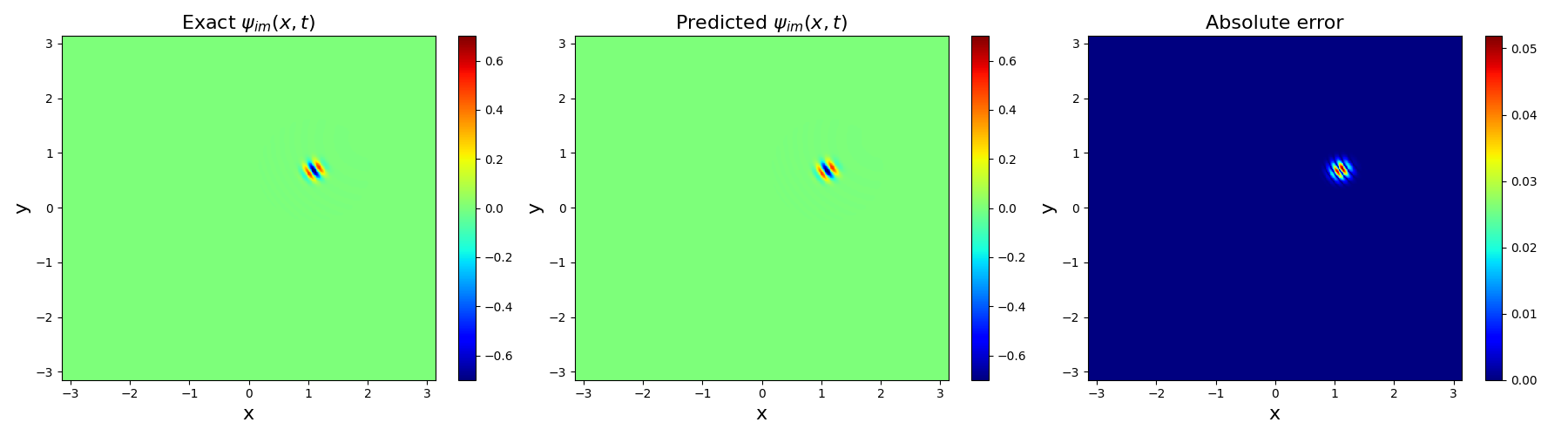}}\hfill
	\caption{2D Schrödinger equation with $\varepsilon=0.01$, harmonic potential $V(\bm{x}) = \frac{x_1^2 + x_2^2}{2}$, and 
	$\bm{q}(0)=(0.8, 0.5)^T, \ \bm{p}(0) = (0.8, 0.5)^T, \ A(0) = \mathrm{i}\text{diag}(0.7, 0.7), \ \gamma(0) = 0.75$.
	(a) Left: the real part of reference solution $\psi_{\text{re}}(\bm{x},1)$. 
	Mid: the real part of prediction solution $\psi_{\text{re}}(\bm{x},1;\theta)$.
	Right: absolute error of real part $\left|\psi_{\text{re}}(\bm{x},1) - \psi_{\text{re}}(\bm{x},1;\theta)\right|$.
	(b) Left: the imaginary part of reference solution $\psi_{\text{im}}(\bm{x},1)$. 
	Mid: the imaginary part of prediction solution $\psi_{\text{im}}(\bm{x},1;\theta)$.
	Right: absolute error of imaginary part $\left|\psi_{\text{im}}(\bm{x},1) - \psi_{\text{im}}(\bm{x},1;\theta)\right|$.}
	\label{op_2d_2}
\end{figure}

To train the networks, $N=2,000$ initial functions defined by (\ref{op_ini}) are taken as 
the training set, with parameters randomly sampled from the specified distributions.
Additionally, in the loss function (\ref{phy_deeponet_loss}), $Q=500$ time points 
$\{t_j\}_{j=1}^Q$ are randomly sampled from the interval $[0,1]$. 
The physics-informed DeepONets with Gaussian 
wave packets are trained by minimizing the
loss function (\ref{phy_deeponet_loss}) over 200,000 iterations. $N_{\text{test}}=100$ 
initial functions (\ref{op_ini}) are sampled from the above distribution to compute the corresponding mean and standard deviation of the relative
$L^2$ error, which are reported as 3.62e-02 and 5.04e-02, respectively.
The numerical results for various input samples 
from the test set are shown in
Figure \ref{op_2d_1} and Figure \ref{op_2d_2}. The relative $L^2$ errors for these cases 
are 2.39e-02 and 3.74e-02, respectively.
These results demonstrate the accuracy of physics-informed DeepONets 
with Gaussian wave packets in solving the 
Schrödinger equation in two-dimensional space. 
To further evaluate the efficiency of our method, we adopt a similar experimental 
setup as in section~\ref{op_tor_1d} and assess the computational time required to solve 
varying numbers of initial value problems, denoted by $N_{\text{test}}$. 
As reported in Table~\ref{ex3_compare_t}, DeepONets consistently outperform 
the classical fourth-order RK4 scheme, achieving a speedup of at least two to three times. 
Notably, the runtime of RK4 also increases approximately linearly with $N_{\text{test}}$, 
whereas the computational cost of DeepONets remain nearly constant. These results further 
demonstrate the superior computational efficiency and scalability of the proposed operator 
learning approach.

\begin{table}[htbp]
  \begin{center}
  \caption{Runtime comparison of RK4 and DeepONets for different 
  $N_{\text{test}}$ initial conditions when $V(\bm{x}) = \frac{x_1^2 + x_2^2}{2}$.}
  \begin{tabular}{cccc}
    \hline
    $N_{\text{test}}$ & $10,000$ & $20,000$ & $40,000$ \\
    \hline 
     RK4 & 7.31s & 14.66s & 29.26s \\
    \hline
     DeepONets & 2.72s & 3.38s & 3.65s \\
    \hline 
  \end{tabular}
  \label{ex3_compare_t}

  \end{center}
\end{table}

\section{Summary}
\label{summary}
In this paper, we investigate the application of neural 
networks to solving the semi-classical limit of the Schrödinger equation.
Due to high-frequency oscillations in both space and time, directly solving this equation using neural 
networks yields low accuracy for small $\varepsilon$. To address this, we employ 
Gaussian wave packets to transform the original problem into a system of ODEs and then solve this ODE system using neural networks, such as PINNs and MscaleDNNs. 
Due to the presence of multiscale properties in this ODE system, MscaleDNNs significantly outperform PINNs in solving it.
Additionally, we apply the physics-informed DeepONets with Gaussian wave packets to map initial values to solutions. Numerical examples highlight the capability of physics-informed DeepONets with Gaussian wave packets
to simultaneously obtain solutions for different initial conditions.\par

Despite these advancements, further theoretical and 
computational research is necessary. In particular, a robust 
theoretical framework explaining why MscaleDNNs outperform 
PINNs in solving ODE systems remains elusive. 
Furthermore, addressing the operator problem for the 
Schrödinger equation may benefit from exploring specialized 
network architectures. 
In addition, it would be natural to extend the Gaussian 
wave packet-based neural framework to broader classes of linear 
Schrödinger models, such as those involving magnetic fields or periodic 
potentials. A particularly promising avenue is the application to multi-state 
systems with nonadiabatic transitions, which would require adapting both the 
Gaussian wave packet formalism and the network design to handle coupled potential 
energy surfaces and vector-valued wave functions.

\appendix
\section{Extension to Hagedorn wave packets}
	\label{ha_wave_packets}
	We consider Hagedorn wave packets \cite{faou2009computing, lasser2020computing} in the one-dimensional setting $(d=1)$, where the solution of equation~\eqref{semi_sch} is represented as

    \begin{equation*}
      \psi(x, t) = \sum_{k=0}^K c_k(t) \phi_k(x, t),
    \end{equation*}
    with $c_k(t)$ are the time-dependent coefficients associated with each basis function $\phi_k(x,t)$. When the potential is harmonic, the coefficients $c_k(t)$ remain constant over time and can be computed directly from the initial condition via projection:
    \begin{equation*}
        c_k = \langle \psi(x,0), \phi_k(x,0) \rangle.
    \end{equation*}
    In contrast, for anharmonic or more general potentials, the coefficients $c_k(t)$ evolve over time and must be determined by solving additional linear ODEs. In the following experiments, we focus on the harmonic case and assume that all coefficients are identically equal to one. This assumption simplifies the representation and allows us to isolate the effect of the basis functions themselves, as the dynamics of the coefficients in more general settings are independent of our neural network approach and can be handled using standard methods.
    
    The basis functions are defined as follows:
    \begin{equation*}
      \phi_0(x, t) = \left(\frac{1}{\pi \varepsilon}\right)^{1/4} \frac{1}{\sqrt{Q(t)}} \cdot \exp\left[\frac{\mathrm{i}}{\varepsilon}
      \left(\frac{P(t)}{2Q(t)}\left(x-q(t)\right)^2 + p(t)(x-q(t)) + S(t)\right)\right],
    \end{equation*}
	and for $k\geq 1$,
    \begin{equation*}
      \phi_k(x, t) = \frac{1}{\sqrt{2^k k!}}\cdot H_k \left(\frac{x - q(t)}{\sqrt{\varepsilon}}\right)\cdot \phi_0(x, t),
    \end{equation*}
    where $q(t), \, p(t), \, S(t) \in \mathbb{R}$, 
    $Q(t),\, P(t)\in \mathbb{C}$, and 
    $H_k$ represents the $k$-th order Hermite polynomial.

	Unlike simple Gaussian wave packets, which are typically described by a single complex-valued function $\alpha(t)$, Hagedorn wave packets are constructed from a structured, time-evolving orthonormal basis determined by $Q(t)$ and $P(t)$. 
	The parameters $q(t), \, p(t), \, Q(t), \, P(t), \, S(t)$ 
	evolve according to the following system of ODEs:
    \begin{equation*}
    \begin{cases}
      \dot{q}(t)&=p(t), \\
      \dot{p}(t)&=-\nabla V(q(t)), \\
      \dot{Q}(t)&= P(t),\\
      \dot{P}(t)&= - \nabla^2 V(q(t))\cdot Q(t), \\
      \dot{S}(t)&= \frac{1}{2}\left|p(t)\right|^2 - V(q(t)).
    \end{cases}
    \end{equation*}
	These equations describe the evolution of the Hagedorn parameters along classical trajectories, capturing both the transport and dispersion characteristics of the wave packet.
	In our work, one may solve a single initial value problem 
	using PINNs or MscaleDNNs. Alternatively, operator learning approaches such as DeepONet can be employed to learn solution operators for a family of initial value problems sharing the same potential $V(x)$.

\subsection{PINNs and MscaleDNNs with Hagedorn wave packets}

    Consider the harmonic potential $V(x)=\frac{1}{2}x^2$, 
	with initial parameters 
	$q(0)=1.0, \, p(0)=2.0, \, Q(0)=1.0, \, P(0)=\mathrm{i}, \, S(0)=1.0$, 
	and terminal time $T=1.0$. 
	For this potential, the coefficients $c_k(t)$ in the Hagedorn 
	wave packet expansion remain constant over time and are 
	determined solely by the initial projection 
	$\langle \psi_0, \phi_k(x,0) \rangle$. 
	For simplicity, we assume unit coefficients and 
	consider initial data concentrated entirely on the first $K=10$ modes. The initial wavefunction is thus defined as

    \begin{equation}
	  \label{ha_ini}
      \psi(x, 0) = \sum_{k=0}^{10} \phi_k(x, 0).
    \end{equation}
    Here, the ground mode $\phi_0(x, 0)$ is given by
    \begin{equation*}
      \phi_0(x, 0) = \left(\frac{1}{\pi \varepsilon}\right)^{1/4} \cdot \exp\left[\frac{\mathrm{i}}{\varepsilon}
      \left(\frac{\mathrm{i}}{2}\left(x-1\right)^2 + 2x - 1\right)\right],
    \end{equation*}
	and the higher-order modes are constructed as
    \begin{equation*}
      \phi_k(x, 0) = \frac{1}{\sqrt{2^k k!}}\cdot H_k \left(\frac{x - 1}{\sqrt{\varepsilon}}\right)\cdot \phi_0(x, 0).
    \end{equation*}

    We evaluate the performance of PINNs and MscaleDNNs 
    for Hagedorn wave packets across a range of $\varepsilon$ values: 
	$\varepsilon = \frac{4}{25}, \ \frac{1}{25}, \ \frac{1}{100}, \ \frac{1}{400}, \ \frac{1}{1600}$, and $ \frac{1}{6400}$. 
	The experimental setup mirrors that of the 
	Gaussian wave packet example 
	presented earlier in this paper. However, 
	due to differences in the underlying variables and the associated ODE system, the neural network architecture is modified to [100, 400, 400, 400, 7]. 
	The corresponding relative $L^2$ errors are 
	reported in Table~\ref{tab4}. MscaleDNNs consistently achieve improvements of one to two orders of magnitude over standard PINNs, 
	highlighting the effectiveness of integrating MscaleDNNs 
	with Hagedorn wave packets. 
    Moreover, the trend in error variation across different 
$\varepsilon$ values closely resembles that observed in the Gaussian wave packet case.

	\begin{table}[H]
	\centering
	\caption{Relative $L^2$ error of 1D examples when employing PINNs or MscaleDNNs with Hagedorn wave packets for the Schrödinger equation (\ref{semi_sch})
	with $\varepsilon = \frac{4}{25}, \ \frac{1}{25}, \ \frac{1}{100}, \ \frac{1}{400}, \ \frac{1}{1600}, \ \frac{1}{6400}$, and $V(x) = \frac{1}{2}x^2$.}
	\label{tab4}
	\begin{tabular}{ccccccc}
	\hline
	$\varepsilon$ & $\frac{4}{25}$ & $\frac{1}{25}$ & $\frac{1}{100}$& $\frac{1}{400}$ & $\frac{1}{1600}$ & $\frac{1}{6400}$ \\
	\hline 
	PINNs & 2.239e-03 & 7.640e-03 & 2.733e-02 & 1.023e-01 & 3.925e-01 &1.306e+00 \\
	\hline
	MscaleDNNs & 1.406e-04 & 3.222e-04 & 1.009e-03 & 3.867e-03 & 1.506e-02 & 5.998e-02\\
	\hline
	\end{tabular}
	\end{table}

\subsection{DeepONets with Hagedorn wave packets}

    We now demonstrate the effectiveness of 
    physics-informed DeepONets in conjunction with Hagedorn wave packets. 
    Consider the harmonic potential $V(x)=\frac{1}{2}x^2$ with $\varepsilon=0.01$ and 
    assume $T=1$. The initial parameters are sampled as follows:
    \begin{equation*}
      q(0),\, Q_{\text{re}}(0), \, P_{\text{im}}(0), \, S(0) \sim \mathcal{U}[0.5, 1.5], \quad p(0) \sim \mathcal{U}[1.5, 2.5],
    \end{equation*}
    where $Q_{\text{im}}(0)=0, \, P_{\text{re}}(0)=0$, and 
    $\mathcal{U}$ denotes a uniform distribution. 
    We represent the solution operator using DeepONets
    $\mathcal{G}_{\theta}$, as defined in 
    formula (\ref{multi_deeponet}), with 
    $I=7$ and $J_1 = \dots = J_I = 100$.
    The branch network architecture is [7, 100, 100, 100, 100, 700],
    and the trunk network architecture is [1, 100, 100, 100, 100, 100]. 
    
	We generate $N=2,000$ 
	initial wavefunctions, as defined by equation \eqref{ha_ini}, 
	using random samples of $q(0)$, $p(0)$, $Q_{\text{re}}(0)$, 
    $P_{\text{im}}(0)$, and $S(0)$ drawn from the specified distributions. 
	The remaining experimental setup closely follows that of the Gaussian wave packet example presented earlier in the paper. To assess model performance, we evaluate the relative 
    $L^2$
  error over a test set of 
    $N_{\text{test}}=100$ initial functions drawn from the same distribution. The mean and standard 
	deviation of the relative $L^2$ error for the final solution $\psi$ are $\text{mean}(E_{\text{rel}})=$2.18e-02  
	and $ \text{std}(E_{\text{rel}})$=2.35e-02, respectively. 
	The corresponding statistics for the physical quantities
    $q,\, p,\, Q,\, P,\, S$, and the
    final solution $\psi$ 
    are summarized in Table \ref{tab5}.
    Notably, for $\varepsilon=0.01$, 
	both the mean and standard deviation of the relative 
	$L^2$ error in the final solution are approximately two 
	orders of magnitude larger than those of the corresponding 
	ODE system solution, consistent with the results observed in the Gaussian wave packet case.

	\begin{table}[H]
	\centering
	\caption{The mean and standard deviation of relative $L^2$ 
	error for $q$, $p$, $Q$, $P$, $S$, 
	and the final solution $\psi$.}
	\label{tab5}
	\begin{tabular}{ccccccc}
	\hline
	& $q$ & $p$ & $Q$ & $P$ & $S$ & $\psi$ \\
	\hline 
	mean & 9.71e-05 & 1.99e-04 & 1.33e-04 & 1.29e-04 & 2.98e-04 & 2.18e-02 \\
	\hline
	standard deviation & 9.06e-05 & 2.47e-04 & 1.96e-04 & 1.93e-04 & 4.00e-04 & 2.35e-02 \\
	\hline 
	\end{tabular}
	
	\end{table}

\section{Gaussian beam decomposition for WKB-type initial condition}
\label{gbd_wkb}
When applying Gaussian beam decomposition to solve the 
semi-classical limit of the Schr{\"o}dinger equation with 
WKB-type initial data, one typically encounters a large 
number of decoupled ODE systems, often with complexity 
dependent on the semi-classical parameter $\varepsilon$. 
DeepONets can be effectively integrated with Gaussian beam 
decomposition, as highlighted in section~\ref{op_tor_1d}, 
offering an efficient approach for solving such a large number 
of ODE systems. To illustrate this, we present a 
one-dimensional example on the time interval $t\in[0,1]$, 
with the potential energy function given by $V(x)=\frac{1}{2}x^2$, 
and WKB-type initial data:

\begin{equation*}
	\psi(x, 0) = a(x)\exp\left(\frac{\mathrm{i}}{\varepsilon} S(x)\right),
\end{equation*}
where $\varepsilon = 0.04, \ a(x) = \exp(-25(x-0.5)^2)$ 
and $S(x)= x + 1$.

The solution of the Schr{\"o}dinger equation can be approximated by a superposition of Gaussian 
beams:
\begin{equation*}
	\psi(x, t) \approx \sum_{j=1}^{\mathfrak{N}} A_{j} \exp\left(\frac{\mathrm{i}}{\varepsilon}\Phi_j(x, t) \right),
\end{equation*} 
where $\mathfrak{N}$ denotes the number of Gaussian beams. 
Each beam is centered along a classical trajectory $q_j(t)$, 
and is characterized by a phase function of the form:
\begin{equation*}
	\Phi_j(x, t) = S_j(t) + p_j(t)\left(x - q_j(t)\right) + \frac{1}{2}M_j(t)\left(x - q_j(t)\right)^2.
\end{equation*}
For each $j=1,\dots, \mathfrak{N}$, 
the quantities $q_j(t), \ p_j(t) \in \mathbb{R}$, and 
$\ M_j(t), \ S_j(t)\in \mathbb{C}$ 
satisfy the following system of ODEs:

\begin{equation}
\label{gb_ode}
\begin{cases}
\dot{q}_j = p_j, \\
\dot{p}_j = -V'(q_j), \\
\dot{M}_j = -M_j^2 - V''(q_j), \\
\dot{S}_j = \frac{1}{2}p_j^2 - V(q_j) + \frac{\mathrm{i}}{2} M_j \varepsilon.
\end{cases}
\end{equation}
Then, $q_j(0)$ are sampled from the interval $[-0.1, 1.1]$ 
with uniform spacing, and the corresponding initial 
conditions are set as $p_j(0)=1.0, \ M_j(0)=25\mathrm{i}$, and 
$S_j(0)=S(q_j)$, for $j=1, \dots, \mathfrak{N}$. 
As a result, we need to solve $\mathfrak{N}$ instances of the ODE 
system \eqref{gb_ode}, each with distinct initial conditions. 
To address this efficiently, we construct DeepONets model in a 
manner analogous to section~\ref{pd_gwp}. The loss function 
for the operator network $\mathcal{G}'_{\theta}$ 
is formulated based on the ODE system \eqref{gb_ode}.

In this example, we consider 
$q_j(0) \sim \mathcal{U}[-0.1, 1.1]$, where
$\mathcal{U}$ denotes a uniform distribution. 
This choice allows us to simultaneously 
obtain the final solutions corresponding 
to $q_j(0)$ sampled from $[-0.1, 1.1]$ with uniform spacing. We set $T=1$ 
and specify the initial conditions as: 
\begin{equation*}
	p_j(0) = 1, \ M_j(0) = 25 \mathrm{i}, \ S_j(0) = S(q_j(0)).
\end{equation*}
The objective is to learn the operator that maps the 
initial state $\bm{y}_j(0)=\left(q_j(0), p_j(0), M_j(0), S_j(0)\right)$ 
to the solution $\bm{y}_j(t)$ for $t\in [0,1]$, without relying on 
any paired input-output data. To this end, we represent 
the operator using the DeepONets $\mathcal{G}'_{\theta}$, 
as defined in formula \eqref{multi_deeponet}, with 
$I=6$ and $J_1 = \dots = J_I = 100$. 
The DeepONets consist of a branch network with 
architecture [6, 100, 100, 100, 100, 600] 
and a trunk network with architecture 
[1, 100, 100, 100, 100, 100], 
where the numbers indicate the 
number of neurons in each layer.

We also sample $N=2,000$ initial functions, 
each corresponding to a distinct value of $q_j(0)$ 
uniformly drawn from the interval $[-0.1,1.1]$, 
along with their associated initial conditions 
$p_j(0), \ M_j(0)$ and $ S_j(0)$. Other experimental settings follow 
those described in section~\ref{op_tor_1d}. Once the DeepONets 
model is trained, any Gaussian beam solution can be evaluated 
with nearly constant computational cost. When computing the 
solution with a time step of $\Delta t= 0.01$, the 
results for different values of $\mathfrak{N}$, representing the number 
of Gaussian beams, are presented in Table~\ref{table_gb}.

\begin{table}[htbp]
  \begin{center}
  \caption{Runtime and accuracy of DeepONets when we use different 
  number $\mathfrak{N}$ of Gaussian beams.}
  \begin{tabular}{cccccc}
    \hline
    $\mathfrak{N}$ & $8$ & $32$ & $128$ & $512$ & \\
    \hline 
    Runtime & 3.58s & 3.60s & 3.74s &4.11s \\
    \hline 
	Relative $L^2$ error & 5.29e-01 & 2.95e-02 & 2.95e-02 & 2.95e-02 \\
    \hline
  \end{tabular}
  \label{table_gb}
  \end{center}
\end{table}

As shown in Table~\ref{table_gb}, increasing the number $\mathfrak{N}$ 
of Gaussian beams leads to progressively improved accuracy, 
which eventually stabilizes. Notably, the computation time remains 
nearly constant—approximately 3 seconds—regardless of the value of $\mathfrak{N}$. In contrast, the simulation time of the RK4 scheme increases linearly with the value of $\mathfrak{N}$, making it inefficient for large $\mathfrak{N}$. 
It is also worth highlighting that our method does not require 
interpolation when evaluating solutions at finer temporal 
resolutions. For instance, with a time step of $\Delta t=0.005$ 
and $\mathfrak{N}=128$ Gaussian beams, the total runtime is only 4.47 seconds. These results 
collectively demonstrate the computational efficiency 
of our proposed approach.

\section*{Acknowledgement}
The authors wish to thank the anonymous referees for their
thoughtful comments, which helped in the improvement of the presentation.
\nocite{*}

\bibliography{ref}	


\end{document}